\documentclass[useAMS,usenatbib]{mn2e}
\usepackage{amsmath}
\usepackage{wasysym}
\usepackage{graphicx}
\usepackage{epsfig}
\usepackage{epic}
\usepackage{eepic}
\usepackage{color}
\usepackage[usenames,dvipsnames]{xcolor}

\usepackage{float}

\renewcommand{\vec}[1]{ {\bmath #1} }

\newcommand{\mbi}[1]{\mbox{\boldmath$#1$}}

\newcommand{\lsim}{\mbox{${\,\hbox{\hbox{$ < $}\kern -0.8em \lower 1.0ex\hbox{$\sim$}}\,}$}}
\newcommand{\gsim}{\mbox{${\,\hbox{\hbox{$ > $}\kern -0.8em \lower 1.0ex\hbox{$\sim$}}\,}$}}

\newcommand{\shortminus}{\scalebox{0.5}[1.0]{\( - \)}}

\def\beqn{\vspace{2mm}
\begin{eqnarray}} 
\def\eeqn{\vspace{2mm} 
\end{eqnarray}}

 \def\la{\langle}

\newcommand{\be}{\begin{equation}}
\newcommand{\ee}{\end{equation}}
\newcommand{\ba}{\begin{eqnarray}}
\newcommand{\ea}{\end{eqnarray}}
\newcommand{\brr}{\begin{array}}
 
\newcommand{\err}{\end{array}}
\newcommand{\bc}{\begin{center}}
\newcommand{\ec}{\end{center}}

\title[Structure Formation of the Local Universe]{Simulating Structure Formation of the Local Universe}

\author[Steffen He{\ss} et. al.]{Steffen He{\ss}\thanks{E-mail: shess@aip.de},
Francisco-Shu Kitaura\thanks{E-mail: kitaura@aip.de, Karl-Schwarzschild fellow} \& Stefan Gottl{\"o}ber \\
Leibniz-Institut f\"ur Astrophysik Potsdam (AIP), An der Sternwarte 16, D-14482
Potsdam, Germany}

\begin{document}

% \date{Accepted 1988 December 15. Received 1988 December 14; in
% original form 1988 October 11}

% \pagerange{\pageref{firstpage}--\pageref{lastpage}} \pubyear{2002}

\maketitle

\label{firstpage}

\begin{abstract}
In this work we present cosmological $N$-body simulations of the Local Universe
with initial conditions constrained by the Two-Micron Redshift Survey (2MRS)
within a cubic volume of $180\,h^{-1}\,\rm{Mpc}$ side-length centred at the
Local Group.

We use a self-consistent Bayesian based approach to explore the
joint parameter space of primordial density fluctuations and peculiar velocity
fields, which are compatible with the 2MRS galaxy distribution after cosmic
evolution. 
This method (the \textsc{kigen}-code) includes the novel  ALPT (Augmented Lagrangian Perturbation Theory) structure formation model
which combines second order LPT (2LPT) on large
scales with the spherical collapse model on small scales.  Furthermore we describe coherent flows with 2LPT and include a dispersion term to model fingers-of-god
(fogs) arising from virialised structures.
These implementations are crucial to avoid artificial parallel filamentary structures, which appear  when using the structure formation model with the 2LPT approximation and the data with compressed fogs. We assume $\Lambda$CDM cosmology throughout our method.

The recovered initial Gaussian fields are used to perform a set of 25 constrained simulations. Statistically this ensemble of simulations is in agreement with a reference set of 25 simulations based on randomly seeded Gaussian fluctuations in terms of matter statistics, power-spectra and mass functions.

Considering the entire volume of $(180\,h^{-1}\,\rm{Mpc})^3$ we obtain correlation
coefficients of about $98.3\%$ for the cell-to-cell comparison between the
simulated density fields and the galaxy density field in log-space with Gaussian
smoothing scales of $r_{\rm S}=3.5\,h^{-1}\,\rm{Mpc}$ ($74\%$ for $r_{\rm
S}=1.4\,h^{-1}\,\rm{Mpc}$).
% The cross power-spectra show correlations with the
% galaxy distribution down to scales of $2.2-3.0\,h^{-1}\,\rm{Mpc}$
The cross power-spectra show correlations with the galaxy distribution which
weakens towards smaller length scales until they vanish at scales of
$2.2-3.0\,h^{-1}\,\rm{Mpc}$.

The simulations we present provide a fully nonlinear density and velocity field
with a high level of correlation with the observed galaxy distribution  at scales of a few $\rm{Mpc}$. 

\end{abstract}


\sloppy
\section{Introduction}
\label{sec:intro}
The leading paradigm in cosmology is given by hierarchical structure formation, in which the large-scale structure we observe today formed through
gravitationally driven subsequent merging of objects of increasing
size \citep[see][]{WhiteRees1978,Fry1978}.

Cosmological simulations have played a major role in confirming
this theory, obtaining structures statistically similar to the observed ones
(see e.~g.~\citet[][]{Davis-85, Springel05} or for a recent review \citet[][]{Kuhlen2012}).
These studies use random seeded fluctuations permitting one only to compare theoretical models to observations based on statistical properties  \citep[see e.~g. the recent work studying the likelihood of large structures][]{Park2012}.

Such an approach relies on the cosmological principle implying that any sufficiently large
observed region can be taken as a representative part of the Universe. Hence
large surveyed volumes in which the Universe can be considered homogeneous
 are required to suppress the so-called
``cosmic variance''.
  Large volume simulations are especially necessary to study the cosmic
 evolution of various signatures imprinted in the primordial density
 fluctuations,  such as baryon acoustic oscillations (BAOs) \citep[as in][]{KimHorizon2009,Prada2012,AnguloXXL2012,DeusSimulation2012}. % or the integrated Sachs-Wolfe (ISW) effect cite???. 
However, numerical $N$-body simulations become extremely expensive, when in addition to large volumes the
surveying of faint objects is required (i.~e.~high resolutions).

A complementary approach to the ``cosmic variance'' problem consists of trying
to reproduce observed structures in the Local Universe. This would allow for a
direct comparison between theory and observations. In this case smaller volumes
are sufficient, since the particular cosmic realisation of the Universe is taken into account.

{
The concept of this approach is to reconstruct the initial density
fields compatible with the galaxy distribution given a particular structure formation model. 
Cosmological $N$-body simulations using the phase information corresponding to
these fields are expected to {\it naturally} reproduce the observed structures
in the Local Universe. The quality of the phase information in the initial
density fields determines the accuracy with which structures are reproduced by a
cosmological $N$-body simulation.  These kind of
simulations are usually called ``constrained simulations'' (CS).
%One should note, that only the phases at the initial conditions are
% constrained.  
Various methods have been used to obtain Gaussian fields given a set of
constraints (see the seminal works by
\citet[][]{1987ApJ...323L.103B,hoffman,rien} and the more recent ones
\citet[][]{kitaura,kitaura_log,jasche_hamil,kitaura_lyman}). }

Hitherto, a number of pioneering attempts have followed this approach showing the large number
of difficulties one encounters when trying to perform constrained simulations
\citep[see][]{Kolatt1996,Bistolas1998,Mathis2002,Klypin2003,Lavaux2010}. These
challenges range from redshift-space distortions, shot-noise, galaxy bias, to
finding the mapping between Eulerian and Lagrangian space.

Constrained simulations of the local universe have been already extensively used
as a numerical near-field cosmological laboratory \citep[for an overview
see][]{Gottloeber2010}. The predictions of Warm and Cold Dark Matter
cosmological models have been compared with the observed distribution of dwarfs
in the Local Universe \citep[][]{Zavala2009,Tikhonov2009} and the  reionisation
history of the Local Group has been studied \citep[][]{Iliev2011}  Zoomed high
resolution simulations of the Local Group have been used to study the satellites
of the Local Group galaxies \citep[see
e.~g.~][]{Knebe2011A,Libeskind2011,Knebe2011B,diCinto2011,diCinto2012a}, the
formation of the Local Group \citep[][]{Jaime2011}  or recently the stripping of
gas from dwarfs moving through the Local Web \citep[see][]{Benitez2013}.

The methods applied so far, need however a deep revision considering the recent
theoretical advances and the acquisition of new data tracing the Local Universe.
While the displacement field has been neglected in the works using velocity data
\citep[see e.~g.~][]{Klypin2003}, which are
unbiased real-space tracers \citep[see][]{1990ApJ...364..370B,1990ApJ...364..349D,Zaroubi1999,courtois}, displacements have been considered within linear Lagrangian perturbation theory (\citet[][]{1970A&A.....5...84Z} approximation) in most of the works using
galaxy redshifts \citep[][]{Kolatt1996,Mathis2002}. 
 Nevertheless, treating in the
 latter case coherent redshift-space distortions in a very approximate way and thus leading to squashed
 structures perpendicular to the line of sight, as is apparent in the work of
\citet[][]{Mathis2002}. A more sophisticated treatment has been employed in
\citet[][]{Lavaux2010} within
 the Zeldovich approximation, however lacking an analysis of the real-space
 reconstruction.	
{ An improvement may be achieved within this approximation from velocity data
as well, as shown in \citet[][]{Doumler3}.}
It has also been shown that higher order perturbation theory is significantly more precise than the Zeldovich approximation to estimate both the displacement and the velocity fields (see the pioneering works of \citet[][]{1993ApJ...405..449G,1999MNRAS.308..763M,2001ApJS..136....1N,2003A&A...406..393M,2006MNRAS.365..939M}  and more recent works on this subject in \citet[][]{kitaura_lin,kitaura_vel,Neyrinck2012,Kitaura2012ALPT}).

Previous works  used {\it inverse} approaches, in which either
the smoothed density field \citep[][]{Nusser1992} or the
galaxies \citep[][]{Peebles89,NB00,BEN02,Brenier2003,Mohayaee2003} are
moved back in time according to the displacement field initially evaluated in
Eulerian space, hereby effectively assuming a one-to-one relation in the
Eulerian-Lagrangian mapping.
This mapping is degenerate after shell-crossing and a statistical approach becomes necessary.

Alternative  approaches have recently been proposed based on sampling
the posterior distribution function of a Gaussian prior with a likelihood relating the initial fields with the observations through a particular
structure formation model
\citep[see][]{Kitaura2012_ICsCS,jasche12,Wang2013} with its unprecedented application to observations in \citet[][]{Kitaura2012StructureLU}. 

We present in this work for the first time  constrained $N$-body simulations using such a statistical {\it forward} approach  based on observations.
In particular we aim at reproducing the observed large-scale structure as it is 
traced by the distribution of galaxies in the Local Universe assuming
hierarchical structure formation with $\Lambda$CDM cosmology and Gaussian seed
fluctuations.

We have introduced many improvements with respect to previous techniques by using the recently developed Augmented Lagrangian Perturbation Theory (ALPT) structure formation model \citep[see][]{Kitaura2012ALPT},  and modelling redshift-space distortions taking into account the tidal shear tensor and including virialised motions \citep[see][]{kitaura_fogs}.

The quality of the constrained simulations which we present in this work permit
us to reach a new level of precision, in which we can effectively consider them
fully nonlinear reconstructions of the large-scale structure. 
The set of simulations we obtain is highly correlated on large scales
($\gsim98\%$ with $r_{\rm S}=3.5\,h^{-1}\,\rm{Mpc}$), enabling us to study the
influence of large scale modes on statistical processes \citep[in a similar
approach as][]{Aragoncalvo2013A}.
The resemblance to the Local Universe makes it possible to directly compare the
findings of the simulations with the observational data.
The degree in which the observed distribution of matter is reproduced serves to
test our cosmological models and verify our understanding of structure
formation.
A large variety of applications can be done based on these simulations with
unprecedented accuracy, ranging from studies of the nonlinear  density and
peculiar
velocity fields,  over galaxy bias analysis, to study merging histories and dependence on environment of particular objects.

The remainder of the paper is organised as follows. First we introduce the method, then we present the input data followed by our constrained simulations and finally we discuss the results and present the conclusions.

\section[]{Method}
\label{sec:method}

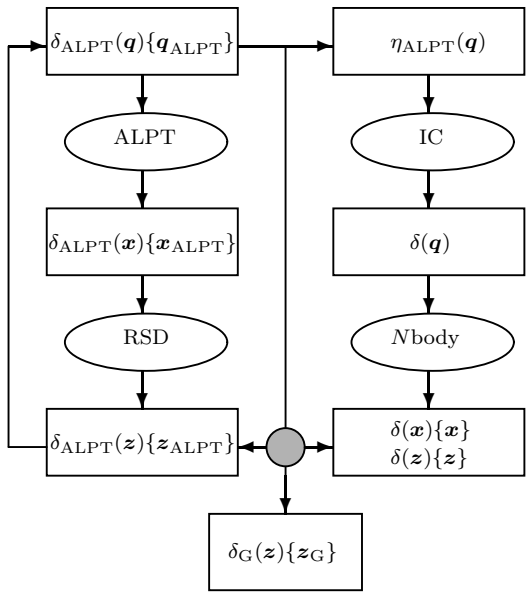
\begin{figure}				
\begin{center}
\vspace{0cm}
\setlength{\unitlength}{0.254mm}
\begin{picture}(337,305)(-20,-310)
        \special{color rgb 0 0 0}\allinethickness{0.254mm}\path(25,-5)(125,-5)(125,-40)(25,-40)(25,-5) % Plain Solid Rectangle
        \special{color rgb 0 0 0}\allinethickness{0.254mm}\put(75,-40){\vector(0,-1){20}} % Vector Line
     %   \special{color rgb 0 0 0}\allinethickness{0.254mm}\put(85,-107){\ellipse{0}{5}} % Plain Solid Ellipse
        \special{color rgb 0 0 0}\put(27,-26){\shortstack{$\delta_{\rm ALPT}(\mbi q)$\{$\mbi q_{\rm ALPT}$\}}} % Plain Text
        \special{color rgb 0 0 0}\put(60,-76){\shortstack{ALPT}} % Plain Text
        \special{color rgb 0 0 0}\allinethickness{0.254mm}\put(77,-75){\ellipse{85}{30}} % Plain Solid Ellipse
        \special{color rgb 0 0 0}\allinethickness{0.254mm}\put(75,-145){\vector(0,-1){20}} % Vector Line
        \special{color rgb 0 0 0}\put(65,-181){\shortstack{RSD}} % Plain Text
        \special{color rgb 0 0 0}\allinethickness{0.254mm}\put(77,-180){\ellipse{85}{30}} % Plain Solid Ellipse
        \special{color rgb 0 0 0}\allinethickness{0.254mm}\path(25,-110)(125,-110)(125,-145)(25,-145)(25,-110) % Plain Solid Rectangle
        \special{color rgb 0 0 0}\put(26,-131){\shortstack{$\delta_{\rm ALPT}(\mbi x)$\{$\mbi x_{\rm ALPT}$\}}} % Plain Text
        \special{color rgb 0 0 0}\allinethickness{0.254mm}\put(75,-90){\vector(0,-1){20}} % Vector Line
%        \special{color rgb 0 0 0}\allinethickness{0.254mm}\put(85,-212){\ellipse{0}{5}} % Plain Solid Ellipse
     	\special{color rgb 0 0 0}\allinethickness{0.254mm}\path(25,-215)(125,-215)(125,-250)(25,-250)(25,-215) % Plain Solid Rectangle
        \special{color rgb 0 0 0}\put(26.8,-236){\shortstack{$\delta_{\rm ALPT}(\mbi z)$\{$\mbi z_{\rm ALPT}$\}}} % Plain Text
        \special{color rgb 0 0 0}\allinethickness{0.254mm}\put(75,-195){\vector(0,-1){20}} % Vector Line
        \special{color rgb 0 0 0}\allinethickness{0.254mm}\path(175,-5)(275,-5)(275,-40)(175,-40)(175,-5) % Plain Solid Rectangle
        \special{color rgb 0 0 0}\allinethickness{0.254mm}\put(225,-40){\vector(0,-1){20}} % Vector Line
%        \special{color rgb 0 0 0}\allinethickness{0.254mm}\put(235,-107){\ellipse{0}{5}} % Plain Solid Ellipse
        \special{color rgb 0 0 0}\put(205,-26){\shortstack{$\eta_{\rm ALPT}(\mbi q)$}} % Plain Text
        \special{color rgb 0 0 0}\put(220,-76){\shortstack{IC}} % Plain Text
        \special{color rgb 0 0 0}\allinethickness{0.254mm}\put(227,-75){\ellipse{85}{30}} % Plain Solid Ellipse
        \special{color rgb 0 0 0}\allinethickness{0.254mm}\path(175,-110)(275,-110)(275,-145)(175,-145)(175,-110) % Plain Solid Rectangle
        \special{color rgb 0 0 0}\allinethickness{0.254mm}\put(225,-90){\vector(0,-1){20}} % Vector Line
        \special{color rgb 0 0 0}\put(215,-131){\shortstack{$\delta(\mbi q)$}} % Plain Text
%        \special{color rgb 0 0 0}\allinethickness{0.254mm}\put(235,-212){\ellipse{0}{5}} % Plain Solid Ellipse
        \special{color rgb 0 0 0}\put(205,-181){\shortstack{$N$body}} % Plain Text
        \special{color rgb 0 0 0}\allinethickness{0.254mm}\put(227,-180){\ellipse{85}{30}} % Plain Solid Ellipse
        \special{color rgb 0 0 0}\allinethickness{0.254mm}\path(175,-215)(275,-215)(275,-250)(175,-250)(175,-215) % Plain Solid Rectangle
        \special{color rgb 0 0 0}\allinethickness{0.254mm}\put(225,-195){\vector(0,-1){20}} % Vector Line
        \special{color rgb 0 0 0}\allinethickness{0.254mm}\put(225,-145){\vector(0,-1){20}} % Vector Line
        \special{color rgb 0 0 0}\put(120,-293){\shortstack{$\delta_{\rm G}(\mbi z)$\{$\mbi z_{\rm G}$\}}} % Plain Text
        \special{color rgb 0 0 0}\put(205,-229){\shortstack{$\delta(\mbi x)$\{$\mbi x$\}}} % Plain Text
        \special{color rgb 0 0 0}\put(205,-244){\shortstack{$\delta(\mbi z)$\{$\mbi z$\}}} % Plain Text
        \special{color rgb 0 0 0}\allinethickness{0.254mm}\path(5,-235)(5,-25) % Plain Solid Line
        \special{color rgb 0 0 0}\allinethickness{0.254mm}\path(110,-270)(190,-270)(190,-310)(110,-310)(110,-270) % Plain Solid Rectangle
        \special{color rgb 0 0 0}\allinethickness{0.254mm}\path(150,-270)(150,-26) % Plain Solid Line
        \special{color rgb 0 0 0}\allinethickness{0.254mm}\put(125,-25){\vector(1,0){50}} % Vector Line
%        \special{color rgb 0 0 0}\allinethickness{0.254mm}\special{sh 0.3}\put(150,-25){\ellipse{20}{20}} % Shade Solid Circle
        \special{color rgb 0 0 0}\allinethickness{0.254mm}\path(25,-235)(5,-235) % Plain Solid Line
        \special{color rgb 0 0 0}\allinethickness{0.254mm}\put(5,-25){\vector(1,0){20}} % Vector Line
        \special{color rgb 0 0 0}\allinethickness{0.254mm}\put(140,-235){\vector(1,0){35}} % Vector Line
        \special{color rgb 0 0 0}\allinethickness{0.254mm}\put(160,-235){\vector(-1,0){35}} % Vector Line
       \special{color rgb 0 0 0}\allinethickness{0.254mm}\special{sh 0.3}\put(150,-235){\ellipse{20}{20}} % Shade Solid Circle
 %        \special{color rgb 0 0 0}\allinethickness{0.254mm}\put(138,-235){\vector(1,0){3}} % Vector Line
 %       \special{color rgb 0 0 0}\allinethickness{0.254mm}\put(162,-235){\vector(-1,0){3}} % Vector Line
        \special{color rgb 0 0 0}\allinethickness{0.254mm}\put(150,-250){\vector(0,-1){20}} % Vector Line
%        \special{color rgb 0 0 0}\allinethickness{0.254mm}\put(150,-250){\vector(0,1){5}} % Vector Line
        \special{color rgb 0 0 0} % Set color to black again (default font color)
\end{picture}
\caption{\label{fig:method} Methodology applied in this work. The left part of the
flow-chart represents the \textsc{kigen}-code to find the large-scale primordial
fluctuations. In an iterative self-consistent way the density fluctuations
$\delta(\mbi q_{\rm ALPT})$ are sampled at Lagrangian coordinates \{$\mbi q_{\rm
ALPT}$\}. Structure formation is modelled within ALPT leading to the density
field in Eulerian space $\delta(\mbi x_{\rm ALPT})$. Then redshift-space
distortions (RSD) are added yielding $\delta(\mbi z_{\rm ALPT})$. A likelihood
comparison (shaded circle) is done with the galaxy field $\{z_{\rm G}\}$. This
process is iterated within a Gibbs- and Hamiltonian sampling scheme. The right
part of the flow-chart shows the selection of samples according to their
cross-correlation with the galaxy field and the subsequent generation of initial
conditions (IC) for the $N$-body runs from the white-noise fields $\eta_{\rm
ALPT}(\mbi q)$. The results ($\delta(\mbi x)$,$\delta(\mbi z)$)  are
analysed and compared with the observations (shaded circle) to gain insights
about our structure formation modelling and improve our reconstructions. }
\subsection
\end{center}
\end{figure}

Our method consists of exploring in a self-consistent Bayesian based approach
the Gaussian primordial fluctuations which after nonlinear cosmic evolution  are
compatible with the galaxy distribution. Here, approximate and efficient gravity
solvers are needed. We then scan the large set of reconstructions to select the
ones which show the largest correlation with the data. The corresponding set of
Gaussian fields is resampled on a finer mesh to generate the initial conditions,
which we finally use to perform high resolution constrained $N$-body
simulations. These  are employed to further confine the models used in the
iterative scheme.  Fig.~\ref{fig:method} shows a flow-chart summarising the
different steps in the method.

{Assumptions}

In our approach we assume that the primordial density fields are Gaussian
distributed.
Hence the two-point correlation function fully characterises the statistics of
the initial conditions. The corresponding power-spectrum is derived with
\textsc{camb} \citep[][]{Lewis:1999bs} using the WMAP7+BAO+h0 parameters
\citep[][]{Komatsu2011}(for definiteness $\Omega_{\Lambda} = 0.728$, $\Omega_{Matter} = 0.272$, $\Omega_{Baryon} = 0.046$, $H_0 = 0.704$, and $\sigma_8 = 0.807$ and $n_s = 0.967$)  . This cosmology is also used to evolve the density
fields both in the reconstruction process using the approximate gravity solver
(see \S \ref{sec:sfm}) and in the constrained simulations using an $N$-body
code.
We assume that the density distribution in the Local Universe can be reproduced
within the framework of $\Lambda$CDM cosmology and neglect for the time being
baryonic physics. This will certainly play an important role at scales smaller
than the Jeans length which is out of the scope of this work.
Furthermore we assume periodic boundary conditions. This assumption is not
critical, as the data considered here (see \S \ref{sec:input}) are increasingly
sparse towards the boundaries (the selection function drops to values of about
$10^{-2}-10^{-3}$). While this will reduce the accuracy close to the boundaries,
the inner regions are not affected considering that the correlation function of
the density fluctuations quickly drops at distances larger than about 10
$h^{-1}$\,Mpc. Peculiar velocities on the other hand can be correlated
up to distances of about 100 $h^{-1}$\,Mpc \citep[see e.~g.~][]{pirin_dhalo}.
Therefore, we will miss the external attractors to the volume considered in our
work \citep[see][to add a posteriori the missing tidal component from velocity
data]{2000ASPC..201..215H,courtois}. The details of the different  models used
in the reconstruction process can be found below.

\subsection{Reconstruction of the large-scale primordial fluctuations}
\label{sec:kigen}

We rely on the \textsc{kigen}-code to recover the large-scale primordial
fluctuations  \citep[][]{Kitaura2012_ICsCS}. The method implemented in this code
is based on sampling Gaussian fields constrained by Lagrangian test particles
$\{\mbi q\}$ \citep[here Hamiltonian sampling is employed as developed
in][]{jasche_hamil,kitaura_lyman},  which after cosmic evolution  (see
\ref{sec:sfm}) and once  redshift-space distortions have been included (see
\ref{sec:rsd})  yields a distribution in Eulerian redshift-space $\{\mbi z\}$
compatible with the observed galaxies $\{\mbi z_{\rm G}\}$. The set of
constraints (matter tracers on a grid)  in Lagrangian space $\{\mbi q\}$ is
obtained from the likelihood comparison between the particles in Eulerian
redshift-space with the galaxies, as the information of each particle's
trajectory is known. Both quantities, the density fluctuations and the set of
particles in Lagrangian space, are iteratively obtained within a Gibbs-sampling
scheme \citep[see][and references therein]{
kitaura}. 
Effectively we sample the posterior distribution function of Lagrangian density
fields $\delta(\mbi q)$  given a set of tracers in redshift-space $\{\mbi z_{\rm
G}\}$, a structure formation model  $\mathcal M$  and a peculiar
velocity field $ v$ which emerges from $\mathcal M$: $P(\delta(
q)|\{ z_{\mathrm G}\},\mathcal M, v)$
% and a peculiar velocity field 
% $\mbi v$: $P(\delta(\mbi q)|\{\mbi z_{\rm G}\},\mathcal M,\mbi v)$.
For more details on the method we refer to
\citet[][]{Kitaura2012_ICsCS,kitaura_fogs}.

\subsubsection{Structure formation model}

\label{sec:sfm}

In order to scan the parameter space of primordial fluctuations we need  an
efficient and computationally fast gravity solver. Most approximate structure
formation models are inaccurate when shell-crossing becomes dominant, i.~e.~when
particles massively start crossing each others trajectories. Too severe
shell-crossing can be prevented with the spherical collapse model
\citep[see][]{Neyrinck2012}. However, one needs to take into account the tidal
field component to be accurate on large-scales \citep[see e.~g.~][]{scocci}.  
Therefore, we use the ALPT approach \citep[see][]{Kitaura2012ALPT}.
This method is based on splitting the displacement field into a long- and a
short-range component. %: $\mbi\Psi=\mbi\Psi^{\rm L}+\mbi\Psi^{\rm S}$. 
The long-range component  is computed by second order LPT (2LPT). This
approximation contains a  nonlocal and nonlinear tidal term. The short-range
displacement field is given by a spherical collapse based solution.  A Gaussian
kernel with a smoothing radius of 4  $h^{-1}$\,Mpc  is used to separate between
both regimes.

\subsubsection{Redshift-space distortions treatment}

\label{sec:rsd}

Coherent redshift-space distortions are modelled within 2LPT. However, we
consider two approaches to treat highly nonlinear redshift-space distortions.
\begin{itemize}
\item {\bf CFOG}:  The first being the widely used one in which the
fingers-of-god (fogs) are compressed \citep[see
e.~g.~][]{Erdogdu2004,Tegmark04}. We use a {\it friends-of-friends} algorithm
taking into account the ellipsoidal distribution along the line-of-sight of
groups of galaxies due to virial motions \citep[see][]{Tegmark-04}.

\item {\bf SFOG}:  In the second approach  we apply a novel sampling scheme
\citep[see][]{kitaura_fogs}.
We follow the ansatz proposed in \citet[][]{kitaura} of splitting the velocity
field into a coherent flow $\mbi v^{\rm coh}_r$ and a dispersion term $\mbi
v^\sigma_r$: $\mbi v_r=\mbi v^{\rm coh}_r+\mbi v^\sigma_r$ within the iterative
reconstruction process \citep[see also][]{kitaura_lyman}.
The dispersion term models the velocity distribution in closely virialised
structures producing elongated structures along the line-of-sight. Here we use a
density dependent dispersion and sample the redshift displacement from a
Gaussian distribution \citep[see][]{kitaura_fogs,Kitaura07}.

\end{itemize}

We will compare both approaches against each other in our numerical calculations
section (Sec.~\ref{sec:numerical}).

\subsubsection{Galaxy bias}

The galaxy bias which is effectively modelled within \textsc{kigen} includes a
nonlinear, nonlocal and a scale-dependent component. In the likelihood
comparison step haloes are constructed at the positions of galaxies within the
ALPT approximation with a friends-of-friends scheme \citep[for a review on the
halo-model see][]{cooray-2002-372}.
A similar procedure was suggested for the 2LPT approximation by
\citet[][]{scocci} and applied in \citet[][]{manera12}.  Further, in the
posterior sampling step of Gaussian fields a scale dependent bias is effectively
modelled.
Hence even though matter tracers may may pick up the galaxy bias, the
Hamiltonian sampling process finds an unbiased Gaussian field.

 The over-all shape of the power-spectrum is strongly determined by the prior
correcting for a $k$-dependent bias. We will show in \S \ref{sec:stats}  that
the characteristic features of the power-spectrum come from the particular
matter distribution in the considered volume. Sampling the full shape of the
power-spectrum within the reconstruction process as suggested in
\citet[][]{kitaura,jasche_gibbs,kitaura_lyman} is out of the scope in this
work.

\subsection{Scanning the posterior distribution function: selection of initial
conditions}

\label{sec:sel}

Once the Markov chain has converged it starts producing Gaussian fields sampled
from the posterior probability distribution function (PDF). 
This enables us to compute various estimators like the mean of the posterior PDF
yielding a smoothed conservative field  \citep[see][and \S
\ref{sec:stats}]{kitaura_log,Kitaura2012StructureLU}. We are, however,
interested in samples which have the full power and at the same time can be
considered as representative of the PDF. The mean properties of this sub-set
should coincide with the corresponding mean of the full set having a smaller
variance \citep[see the case of peculiar motions in ][]{Kitaura2012StructureLU}.
 Therefore we select the set of primordial fluctuations which show the highest
correlation with the galaxy  field after cosmic evolution within a particular
structure formation approximation.
Here we choose a spectral analysis to estimate the degree of correlation between
the dark matter (DM) reconstructions $\delta_{\rm DM}$ and the galaxy
overdensity field $\delta_{\rm G}$, i.~e.~the cross power-spectrum:
\be
XP(k)[\delta_{\rm DM},\delta_{\rm G}]\equiv\frac{\langle|\hat{\delta}_{\rm
DM}(\mbi k)\overline{\hat{\delta}_{\rm G}(\mbi k)}|\rangle}{\sqrt{P_{\rm
DM}(k)}\sqrt{P_{\rm G}(k)}}\,,
\ee
where the ensemble brackets denote angular averaging and $P_{\rm DM}(k)$ are
$P_{\rm G}(k)$ the power-spectra corresponding to the fields $\delta_{\rm DM}$
and $\delta_{\rm G}$, respectively. 
The matter overdensity field is defined by
\be
 \delta_{\rm DM} \equiv  \frac{N_{{\rm DM}i}}{\overline{N_{\rm DM}}} - 1\,,
\ee
where $N_{{\rm DM}i}$  denotes the number counts of simulated particles in cell
$i$ gridded with a clouds-in-cell (CIC) scheme, and $\overline{N_{\rm DM}}$ is
the mean number counts computed dividing the total number of particles by the
total number of cells in the considered volume.
We  define the galaxy overdensity by the maximum likelihood of a Poissonian
distribution to compensate for selection function effects $f_{\rm S}$ \citep[see
e.~g.~][]{kitaura_sdss}:
\be
 \delta_{{\rm G}i} \equiv\frac{N_{{\rm G}i}}{f_{{\rm S}i} \overline{N_{\rm G}}}
- 1\,,
\label{eq:deltaG}
\ee
for each cell $i$, with the mean expected number counts being given by:
$\overline{N_{\rm G}}\equiv {\sum_i N_{{\rm G}i}}/{f_{{\rm S}i}}$.
We impose a threshold $XP_{\rm th}$ at a given scale to filter out the tails of
the posterior PDF and focus on the denser probability regions.
 Approaching the grid-resolution at high $k$ the cross power-spectra start to
fluctuate.  Therefore we choose a value of $k_{\rm th}$ far below the Nyquist
frequency ($k_{\rm th}=1$, see \S \ref{sec:comp}):
\be
   {XP}^l_{\rm{ALPT,G}}(k=k_{\rm th})\equiv{XP}^l(k=k_{\rm th})[\delta_{\rm
ALPT},\delta_{\rm G}]  > XP_{\rm th}\,,
\ee
for sample $l$ in the Markov chain.

Cross power-spectra which satisfy this condition usually show higher
correlations over the entire $k$-space region (see \S \ref{sec:comp}). 
To avoid selecting  correlated samples, which are essentially identical, we
discard first and second neighbours taking the largest correlated one among
them:
\be
  {XP}^l_{\rm{ALPT,G}}(k=k_{\rm th})  > {\rm max}({XP}^{l-2 \;
... \; l+2}_{\rm{ALPT,G}}(k=k_{\rm th}))\,.
\label{selection_crit}
\ee

\subsection{Generation of high resolution initial conditions}

From the set of primordial fluctuations we construct the corresponding
white-noise fields: 
\be
\eta^l(\mbi k)\equiv\frac{\hat{\delta}_q^l(\mbi k)}{\sqrt{P(k)}}\,,
\ee
for sample $l$, with $P(k)$ being the theoretical linear power-spectrum and
$\delta_q$ standing for the overdensity field at the initial conditions in
Lagrangian space $q$. 
We note that these fields are essentially uncorrelated from cell-to-cell, except
for some residual deviations from a flat power-spectrum characteristic of the
particular cosmic realisation.  We checked that dividing $\hat{\delta}_q^l(\mbi
k)$ by the square root of the actual  power-spectrum measured from the density
field itself, i.~e.~with perfectly flat white-noise fields, leads to
significantly less correlated density fields with the galaxy field after
simulating structure formation.
The construction of the white-noise field permits us to increase the resolution
 augmenting the smaller scales on a finer mesh randomly sampling from a Gaussian
with unity variance. The resulting fields can  then be multiplied in
Fourier-space with the linear power-spectrum including modes up to the resampled
scales. 
Here we assume the same cosmology as in the reconstruction algorithm to maximise
the precision of the constrained simulations. From the resulting density
fluctuations one needs to compute the displacement and velocities for the number
of particles required in the $N$-body simulation.
These steps are done within the initial conditions (ICs) generation code
\textsc{ginnungagap}\footnote{publicly available at
https://code.google.com/p/ginnungagap/}, which is fully parallelised and has
been extensively tested within the Zeldovich approximation.

\subsection{Constrained $N$-body simulations}

The initial conditions computed as described in the previous sub-section  are
used to simulate structure formation with a $N$-body code. To avoid
transients
which could originate from the Zeldovich approximation in the generation of the
ICs, we start the $N$-body simulations at a relatively high redshift of
$z=100$. 
 We use \textsc{gadget-3}, an improved version of the publicly available
cosmological code \textsc{gadget-2} \cite[last described in][]{gadget2}. The
code is parallelised for distributed memory machines and uses a hierarchical
tree algorithm for the calculation of gravity. 
We will restrict ourselves in this work to dark matter only $N$-body
simulations and leave a detailed description of the star-forming, cooling and
feedback processes for future work.
The advantage of this step is that we can accurately model the nonlinear regime
 and get a realistic distribution of haloes. Haloes are defined in this context
as spherical regions with a density higher than $\rho>200\times\rho_{crit}$,
where $\rho_{crit}$ is the critical density. To identify these regions we use
the 
 \textsc{ahf}-code\footnote{publicly available http://popia.ft.uam.es/AHF}
\citep[][]{Knollmann2009}. The halo distribution can be used to further test the
reconstruction algorithm and tune the models as we will show below (see \S
\ref{sec:kigen}  and \S \ref{sec:RSD}).

\section{Input data}
\label{sec:input}

Our input data in this work is the 2MASS redshift survey (2MRS), ${\rm
K_s}=11.75$ catalogue, 
 with a median redshift of $z\approx0.028$
\citep[][]{Huchra2012}.
This catalogue provides an almost full sky observation of galaxy
redshifts with a sky coverage of 91\% and uniform
completeness of 97.6\% only limited 
by the presence of the foreground stars 
near the Galactic plane ({\it the Zone of Avoidance}
(ZoA), where $|b|<5^\circ$ and $|b|<8^\circ$ near the Galactic centre).
 The same version of the augmented 2MRS catalogue was used as in
\citet[][]{Kitaura2012StructureLU}, which  was filled in the ZoA with random
galaxies generated from the corresponding longitude/distance bins in the
adjacent strips \citep[][]{yahil}. This method is robust for the width of the
2MRS mask  and has been thoroughly tested \citep[for details see][and references
therein]{pirin_dipole}.

\begin{figure}
\begin{center}
\vspace{8.cm}
\hspace{-7mm}
\includegraphics[width=0.35\textwidth]{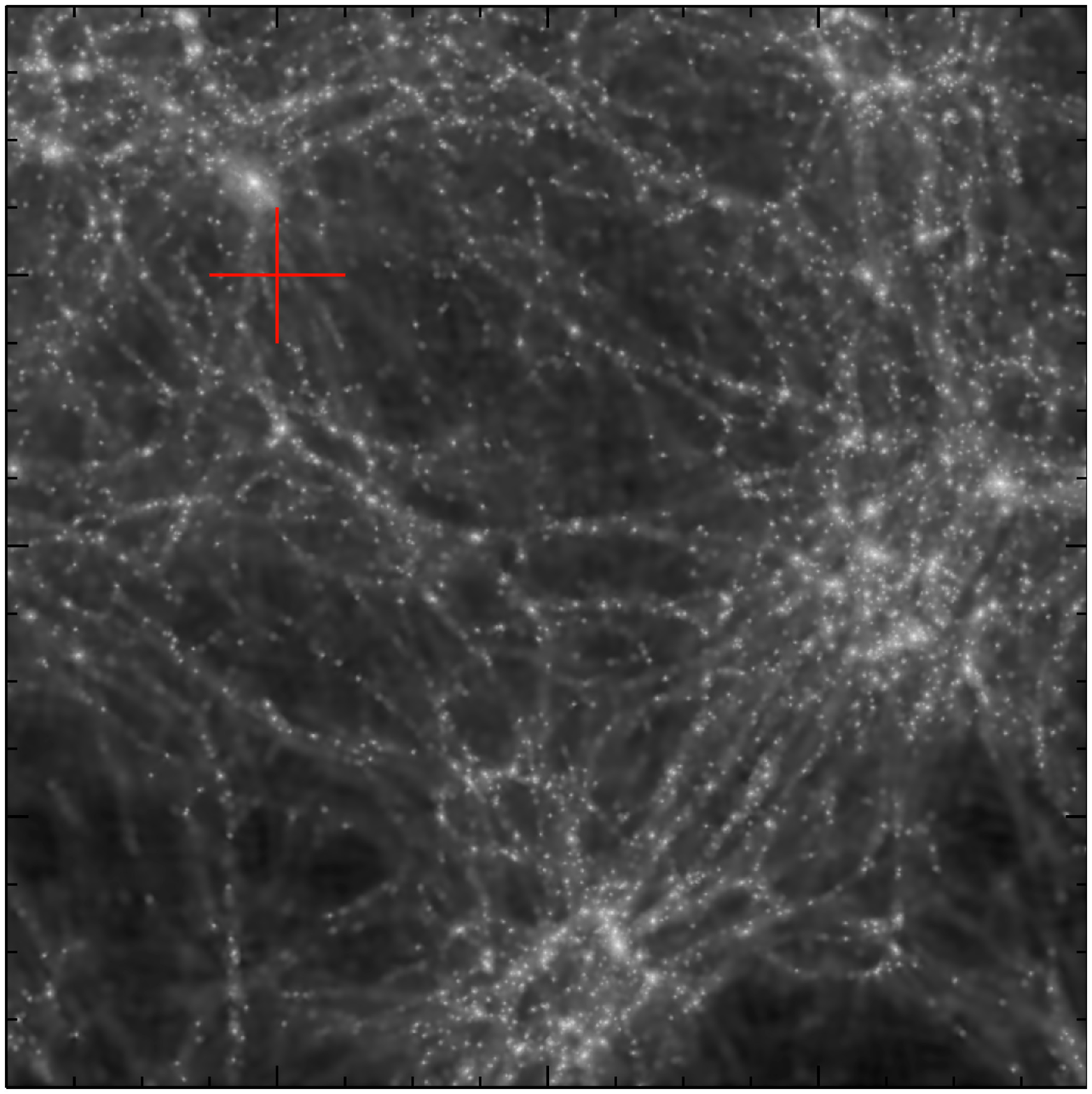}
\put(-159,110){$0$}
\put(-167,73.5){$\shortminus20$}
\put(-167,37){$\shortminus40$}
\put(8,110){$0$}
\put(4,73.5){$\shortminus20$}
\put(4,37){$\shortminus40$}
\allinethickness{0.254mm}\put(-65,22){\color{yellow}\rotatebox[]{-63}{\ellipse{20}{5}}} % Plain Solid Ellipse
\allinethickness{0.254mm}\put(-114.2,125){\color{cyan}\rotatebox[]{-40}{\ellipse{5}{5}}} % Plain Solid Ellipse
\allinethickness{0.254mm}\put(-47,40){\color{green}\rotatebox[]{-40}{\ellipse{26}{45}}} % Plain Solid Ellipse
\put(-180,70){\rotatebox[]{90}{SG Y [$h^{-1}$ Mpc]}}
%\put(24,70){\rotatebox[]{-90}{SG Y [$h^{-1}$ Mpc]}}
%\put(-167,2){$-60$}
\\
\vspace{-0.5mm}
\hspace{-6mm}
\includegraphics[width=0.35\textwidth]{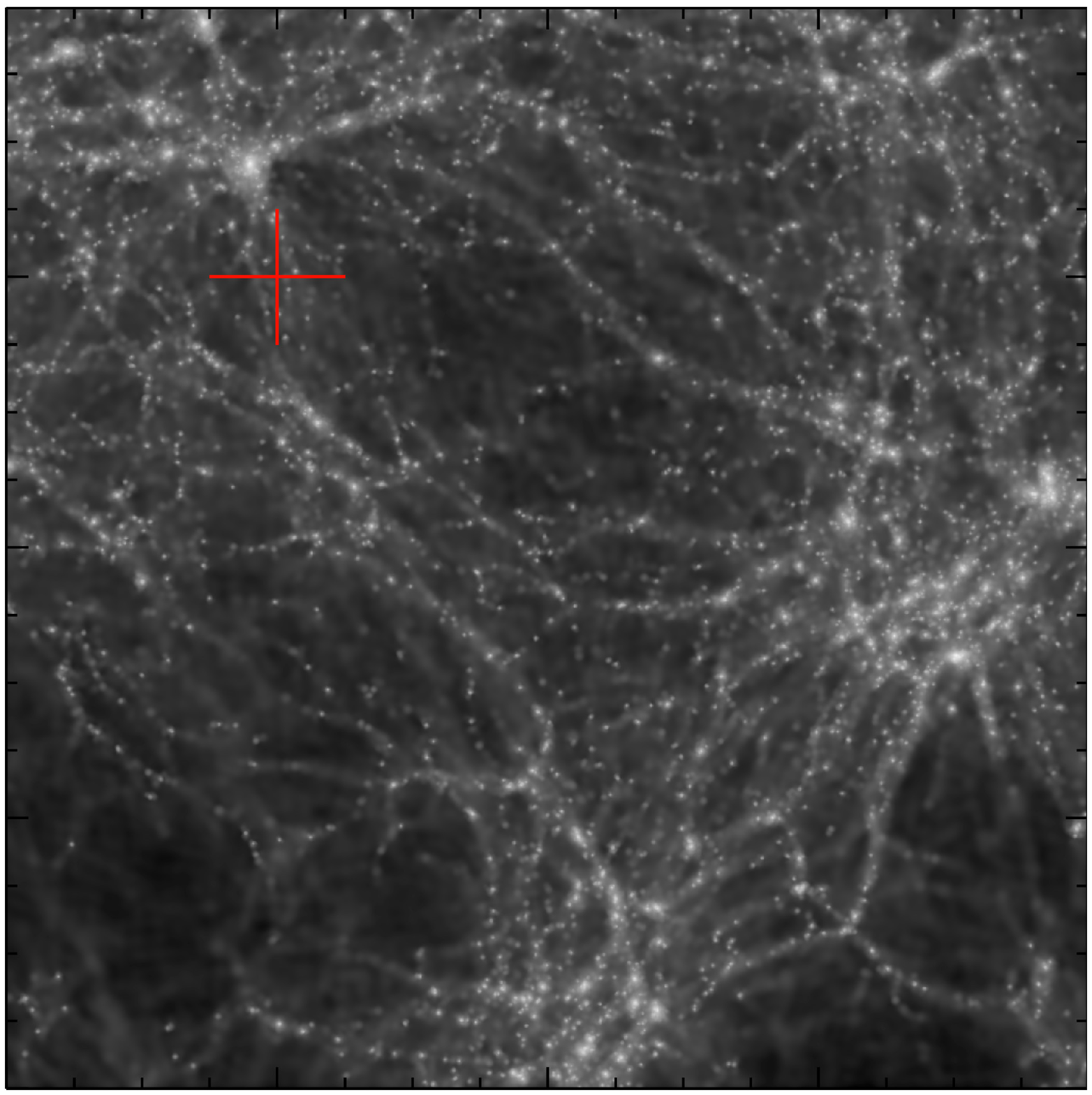}
\allinethickness{0.254mm}\put(-117.36,153){\color{cyan}\rotatebox[]{-40}{\ellipse{5}{5}}} % Plain Solid Ellipse
\put(-183,95){\rotatebox[]{90}{SG Y [$h^{-1}$ Mpc]}}
%\put(21,95){\rotatebox[]{-90}{SG Y [$h^{-1}$ Mpc]}}
\put(-110,6){\rotatebox[]{0}{SG X [$h^{-1}$ Mpc]}}
\put(5,135){$0$}
\put(1,99){$\shortminus20$}
\put(1,63){$\shortminus40$}
\put(-162,135){$0$}
\put(-170,99){$\shortminus20$}
\put(-170,63){$\shortminus40$}
\put(-160,17){$\shortminus20$}
\put(-116,17){$0$}
\put(-81,17){$20$}
\put(-43,17){$40$}
\\
\vspace{-17.68cm}
\hspace{-6mm}
\includegraphics[width=0.35\textwidth]{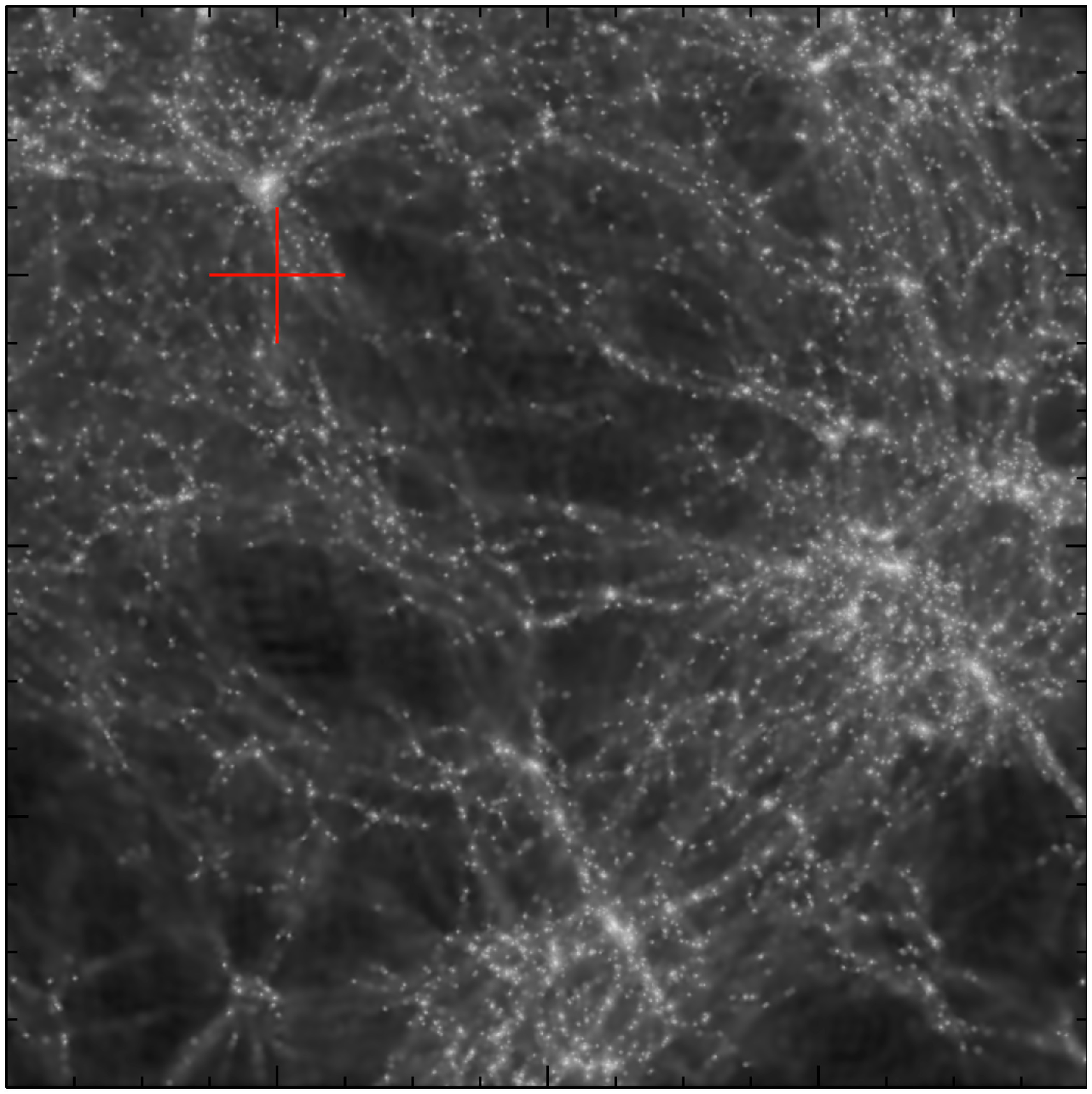}
\allinethickness{0.254mm}\put(-115.7,125.5){\color{cyan}\rotatebox[]{-40}{\ellipse{5}{5}}} % Plain Solid Ellipse
\put(5,110){$0$}
\put(1,73.5){$\shortminus20$}
\put(1,37){$\shortminus40$}
\put(-183,70){\rotatebox[]{90}{SG Y [$h^{-1}$ Mpc]}}
%\put(21,70){\rotatebox[]{-90}{SG Y [$h^{-1}$ Mpc]}}
\put(-162,110){$0$}
\put(-170,73.5){$\shortminus20$}
\put(-170,37){$\shortminus40$}
\allinethickness{0.254mm}\put(-68,26){\color{yellow}\rotatebox[]{-63}{\ellipse{50}{5}}} % Plain Solid Ellipse
\allinethickness{0.254mm}\put(-45,42){\color{green}\rotatebox[]{-40}{\ellipse{35}{45}}} % Plain Solid Ellipse
%\special{color rgb 0 0 0}\allinethickness{0.254mm}\put(-32,91){\color{red}\rotatebox[]{-15}{\ellipse{60}{5}}} % Plain Solid Ellipse
\\
\vspace{-9.4cm}
\hspace{0.9cm}
\begin{minipage}{3.cm}{\rotatebox[]{-20}{\includegraphics[width=0.6\textwidth]{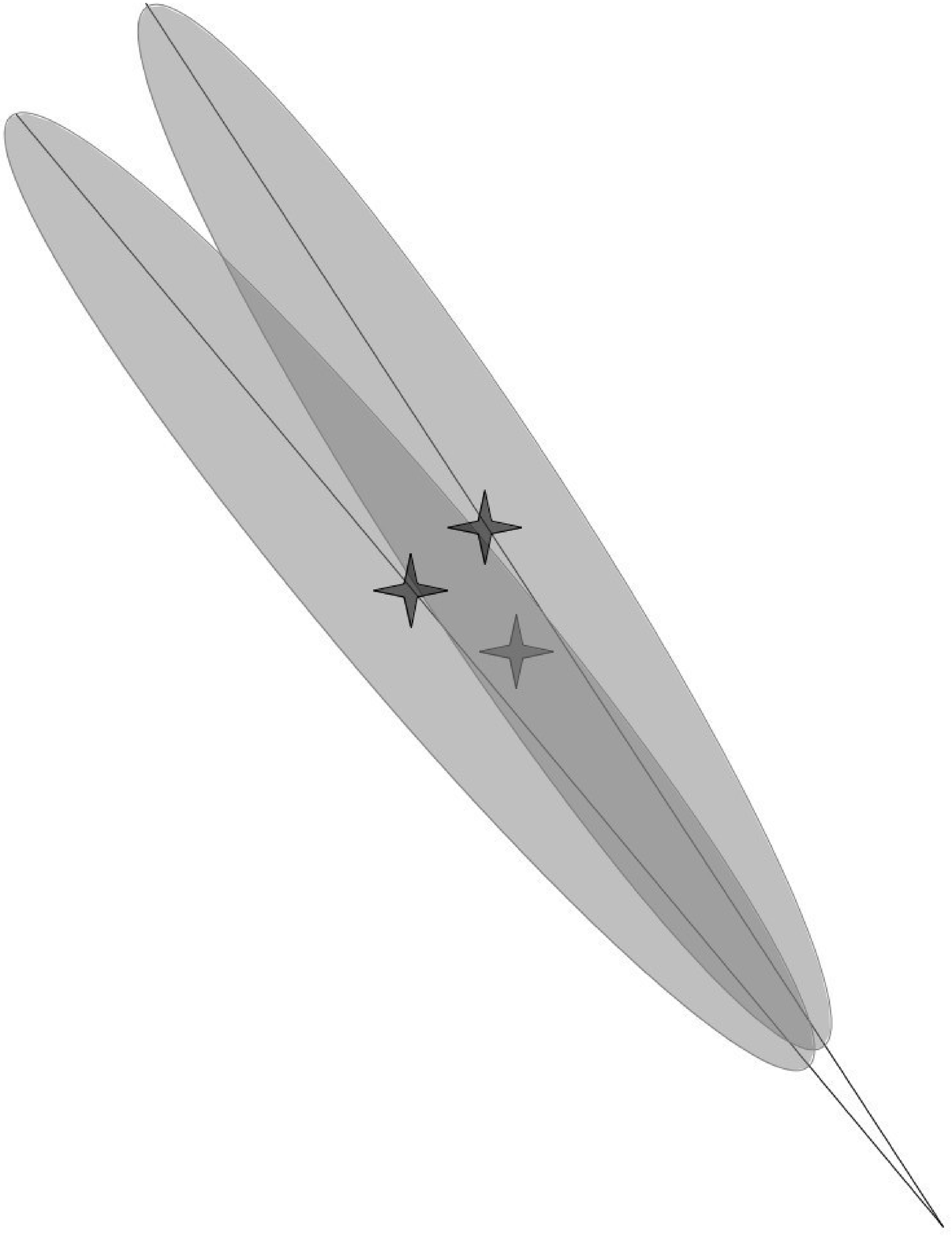}}}
\end{minipage}
\begin{minipage}{4.cm}
\caption{Sketch of overlapping fogs (light grey) close to the observer at the
bottom. A FOF based fog compression might accidentally detect
the darker area and combine the fogs of separated objects (centered at the dark
crosses) to one object closer to the observer (light cross). See the case
of the Virgo cand. (Fig.~\ref{fig:zoom})}
\label{fig:sketch}
\end{minipage}
\vspace{16.4cm}
\caption{Artifacts due to shell-crossing and fog compression (CFOG).
Logarithmic DM overdensity in real-space in a slice of $22.5 \,h^{-1}\,{\rm
Mpc}$ thickness in the supergalactic XY plane. The panels correspond to a
constrained simulation with ({\bf top:}) 2LPT+CFOG, ({\bf middle:})  ALPT+CFOG
and ({\bf bottom:}) ALPT+SFOG. The yellow ellipses show the negative fogs. The
green highlighted area shows the {\it accordion}-effect, a preferential
filament orientation perpendicular to the line-of-sight due to shell-crossing.
The position of the Virgo cluster candidate (cyan circle) is shifted (2-3
$h^{-1}$\,Mpc) towards the observer (red cross) in the upper two panels w.r.t.
the case shown in the lower panel.}
\label{fig:zoom}
\end{center}
\end{figure}

\begin{figure}
\begin{tabular}{cc}
\hspace{-0.75cm}
\includegraphics[width=0.28\textwidth]{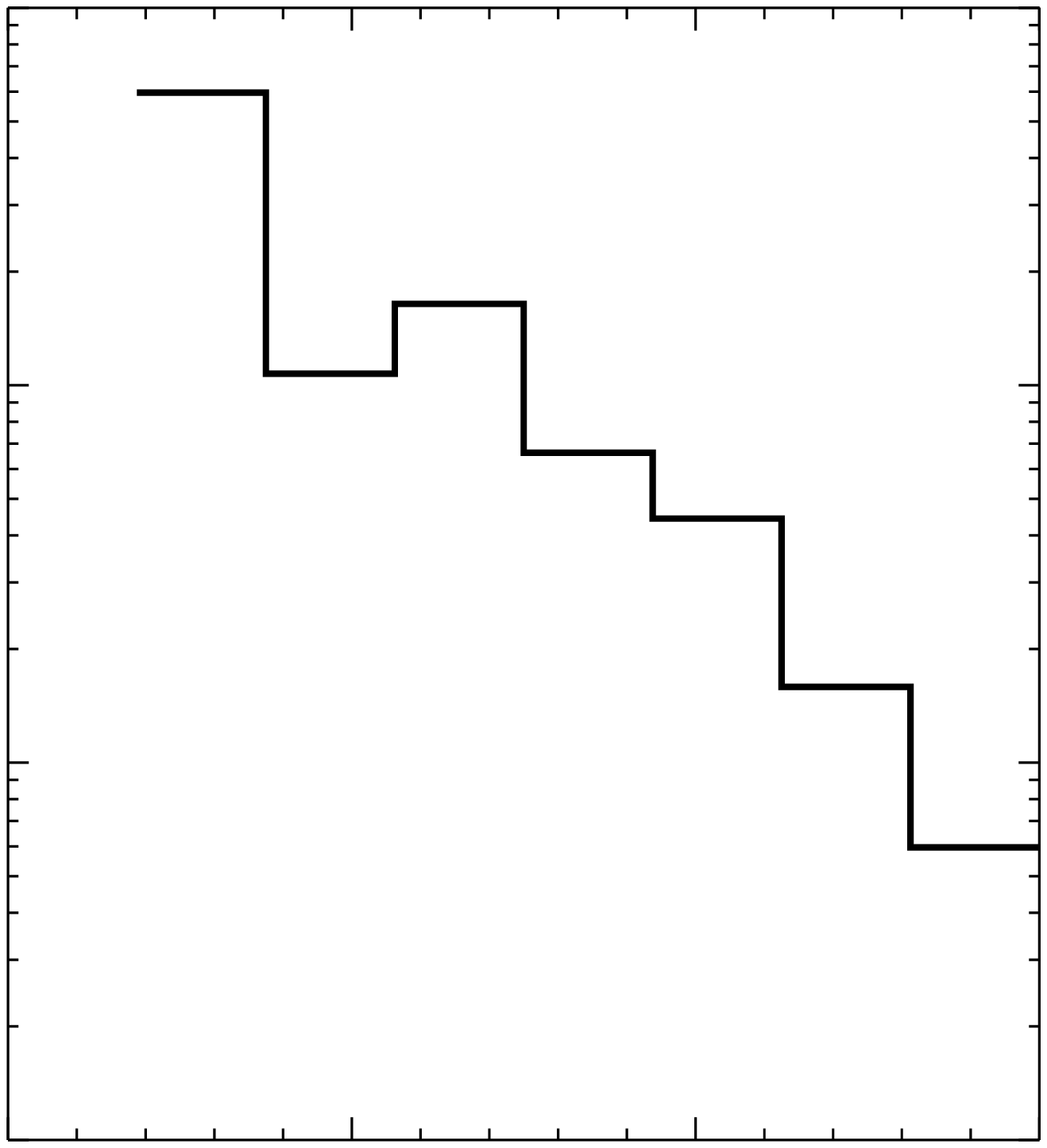}
\put(-120,127){1}
\put(-130,90){$10^{\shortminus1}$}
\put(-130,53){$10^{\shortminus2}$}
\put(-130,17){$10^{\shortminus3}$}
\put(-115,10){$0$}
\put(-83,10){$0.5$}
\put(-49,10){$1.0$}
\put(-15,10){$1.5$}
\put(-140,70){\rotatebox[]{90}{\rm PDF}}
\put(-88,1){$\left| \vec{x}-\langle \vec{x}\rangle\right|\, [h^{-1} {\rm Mpc}]$}
\hspace{-1.1cm}
\includegraphics[width=0.28\textwidth]{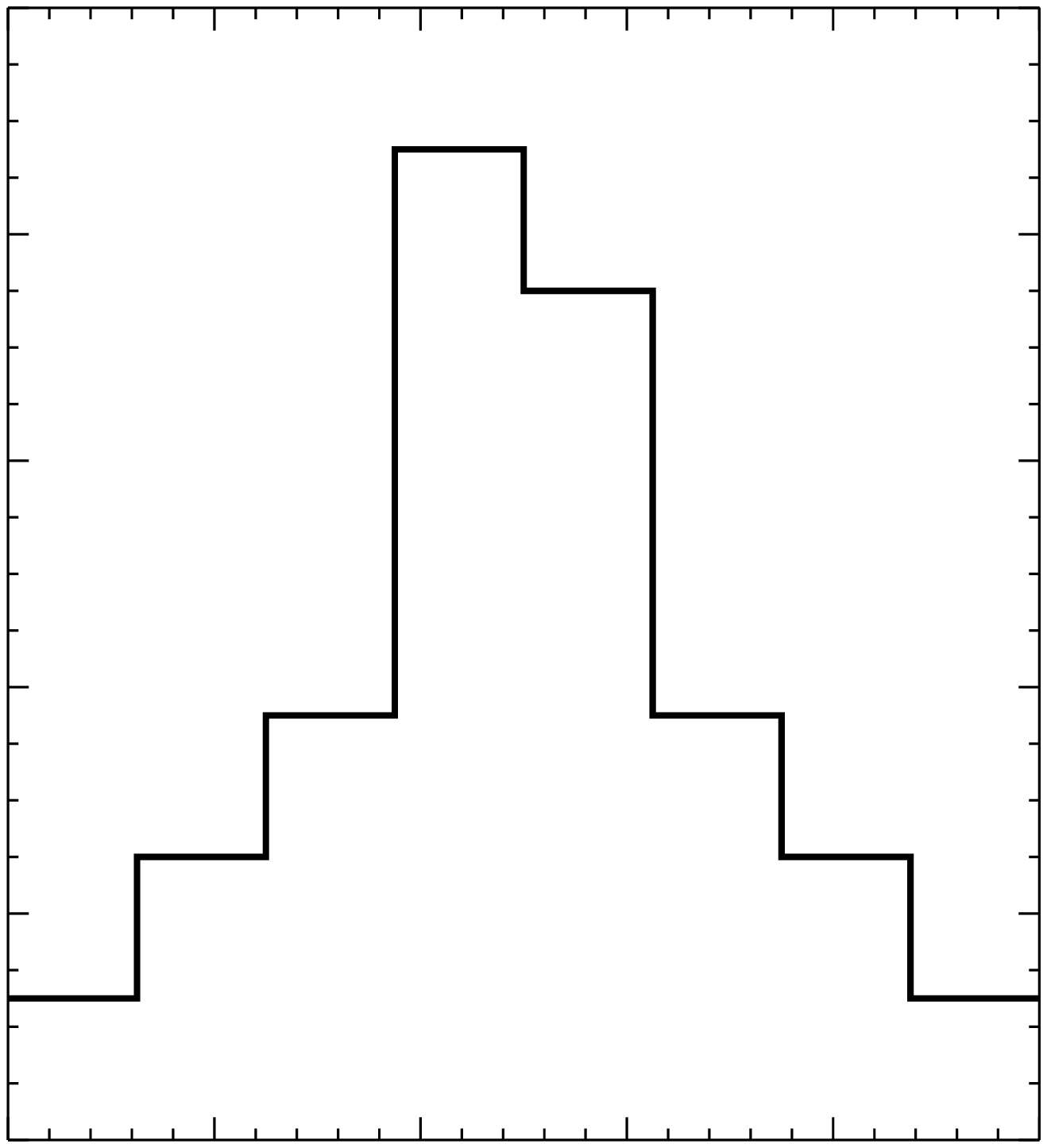}
\put(-5,128){1.0}
\put(-5,107){0.8}
\put(-5,85){0.6}
\put(-5,61){0.4}
\put(-5,39){0.2}
\put(-5,17){0.0}
%\put(-125,128){1.0}
%\put(-125,107){0.8}
%\put(-125,85){0.6}
%\put(-125,61){0.4}
%\put(-125,39){0.2}
%\put(-125,17){0.0}
%\put(-118,10){$1.0$}
\put(-97,10){$1.5$}
\put(-77,10){$2.0$}
\put(-55,10){$2.5$}
\put(-35,10){$3.0$}
\put(-14,10){$3.5$}
\put(-93,1){$M_{200}\, [10 ^{14} h^{-1}M_{\odot}]$}
\end{tabular}
\caption{{\bf Left panel:} PDF of finding the Virgo cluster candidate in
the $25$ constrained simulations (ALPT+SFOG) at a certain distance with respect
to its mean position $\langle \vec{x}\rangle$. {\bf Right panel:} corresponding
mass PDF from the $25$ simulations.}
\label{fig:Virgo}
\end{figure}

 The ZoA could be sampled with a Poissonian likelihood describing the
counts-in-cells of the galaxy distribution which would be limited to a coarse
grid resolution \citep[see e.g.][]{kitaura_log}.
To incorporate a mask treatment within a particle based reconstruction method
like \textsc{kigen},  we would need to sample mock galaxies (or haloes)
according to our structure formation model following schemes like the one
proposed in \citet[][]{scocci}, which is out of scope in this work.  

We computed the selection function $f_{\rm S}$  directly based on the radial
distribution galaxies ignoring the Kaiser-rocket effect
\citep[see][]{2012arXiv1202.5206B}.

A combination of reconstruction and simulation can only reproduce a density
field as good as the data. Hence the stochasticity of galaxy formation and
observational challenges obscure our insight in the true distribution of matter.
Nevertheless due to quantity and precision of position measurements galaxy surveys
are still the most detailed scan of density in the local universe.

\section{Numerical calculations}
\label{sec:numerical}

 In this study we focus on the Local Universe as comprised by a box of
$180 \,h^{-1}\,\rm{Mpc}$ on a side with the observer located in the centre
including  $31017$ 2MRS galaxies (i.~e.~a number density of $5.3\,10^{-3}$
$(h^{-1}\,{\rm Mpc})^{-3}$). 

%\subsection{Large-scale primordial fluctuations}

First we apply the \textsc{kigen}-code to find the large-scale primordial
fluctuations. 
The resolution of the reconstruction is about $1.4\,h^{-1}\rm{Mpc}$ on a mesh of
$128^3$ cells. We follow \citet[][]{Kitaura2012StructureLU} for the numerical
setting of the \textsc{kigen}-code.

The number of constraints ($=$total number of cells $\sim2\,10^{6}$) used in the
Hamiltonian sampling process (see Sec.\ref{sec:kigen})  is achieved by
% taking into account all matter tracers which also trace underdensities.
taking into account all volume elements including empty cells. Therefore we
account for observed regions with no detections. This amount of information
excels that of previous works which relied exclusively on the locations of
density peaks (see Sec.~\ref{sec:intro}) by about 2-3 orders of magnitudes.
Even if one considers that peak positions are less correlated constraints
than gridded volume elements, the latter obviously carries more information.
%  (see e.~g.~\citet[][]{Lavaux2010})
The phase information is used without any additional smoothing to the griding
\citep[see][for smoothed constraints]{Wang2013}. This permits us to recover not
only the density peaks, but also filamentary structures as traced by the galaxy
distribution (see \S \ref{sec:comp}). 

The  power-spectra of the subsequent primordial density fields in the Markov
chain are unbiased with the theoretical linear power-spectrum on all scales as
soon as the \textsc{kigen}-code has converged. For safety we consider samples
after 2000 iterations for our study well after the previous criterion has been
fulfilled. The Markov chain continues running to obtain a minimum of 1000
samples.

Our reconstructions based on 2LPT and the ALPT approximation using data with
collapsed fogs (2LPT+CFOG and ALPT+CFOG, respectively)   lead to constrained
simulations with pronounced artifacts. These are caused by shell-crossing (see
\S \ref{sec:RSD}). As a consequence, the mass function of the case 2LPT-CFOG
clearly deviated from the expected one based on random simulations.
Therefore, we focus in this work on reconstructions using the ALPT structure
formation model with the SFOG (ALPT+SFOG) prescription.
We apply the full machinery sketched in Fig.~\ref{fig:method} to perform
constrained simulations. The simulations are conducted with $384^3$ particles in
a simulation box of $180\,h^{-1}\,\rm{Mpc}$ per side which corresponds to a mass
of $7.8 \times 10^9  \,h^{-1}\,\rm{M_{\odot}}$ of each mass element in the
chosen
cosmology. To this end we used a comoving gravitational softening of
$15\,h^{-1}\,\rm{kpc}$.  

Let us assess the quality of the constrained $N$-body simulations. We will first
analyse our redshift-space distortions (RSD) treatment followed by a comparison
with the observed galaxy field. Finally, we will compare the statistical
properties of the CSs with a set of random simulations.

\subsection{Redshift-space distortions treatment}
\label{sec:RSD}

To  investigate the accuracy of our SFOG treatment we compare it with the
classical CFOG scheme following \citet[]{Tegmark04}. The latter approach shows
significantly larger correlations than the SFOG one. While the CFOG samples
reach cross power-spectrum values of  ${XP}(k=1) \gsim 0.7$, the SFOG ones
remain at  ${XP}(k=1) \lsim 0.5$.  
The artificial collapse of fogs in the CFOG approach leads to relatively more
clustered
regions. This partially compensates for shell-crossing in the final density
fields, and the corresponding lack of power of the approximate
structure formation model.  This effect can induce higher correlations.
% ,however, not necessarily improving the quality of the reconstruction of the
% primordial fluctuations. 
However, a deeper investigation, shows that the CFOG approach induces artifacts.
\input{./fig_xp}

The first one comes from unfolding redshift-space to recover the real-space
matter distribution.  This happens in \textsc{kigen} in a forward approach, by
adding redshift-space distortions to the simulated particles in the likelihood
comparison step \citep[see][]{Kitaura2012_ICsCS}. A correct modelling of RSD is
therefore necessary to obtain accurate reconstructions. One of the problems of
using an approximate structure formation model based on perturbation theory is
that the highly nonlinear virialised motions are not included.  
As shown in \citet[][]{hamilton-1998}  coherent redshift-space distortions can
already produce fogs. Galaxies coherently approaching each other in a potential
well, would produce shell-crossing in redshift-space due  to their peculiar
motions,  leading to elongated structures. Since in the CFOG case \textsc{kigen}
tries to match the compressed data, negative fogs appear in real-space. This
effect is boosted with artificial shell-crossing, as can be seen when using the
2LPT approximation (see fogs in real-space in the upper two panels of
Fig.~\ref{fig:zoom}, where we zoomed into an area
of $80\,h^{-1}\,\rm{Mpc}$).
Treating only coherent flows also fans out  parallel filamentary  structures
like an {\it accordion}, as is apparent in the green highlighted areas in
Fig.~\ref{fig:zoom}. One can see in the extension of the  parallel filamentary
structures that shell-crossing is more severe with 2LPT than ALPT (compare green
areas).
  Another important aspect is the uncertainty in the location of the mass
distribution in high density regions. While the CFOG scheme ignores this point
yielding a single solution, the SFOG sampling approach propagates the
uncertainty in each Markov chain sample. A fogs compressed based reconstruction
is therefore underestimating the uncertainty in high density regions.  
Moreover, the CFOG compression produces systematic artificial displacements of
clusters located close to the observer. This situation is depicted in
Fig.~\ref{fig:sketch}. In regions where various fogs overlap it is difficult to
find the centres of mass with friends-of-friends algorithm. We found that
clusters in such fogs are  systematically located  closer to the observer.
Galaxies residing in the distant tails of fogs, which are more separated, may
not be considered {\it friends} and hence a single shorter fog is detected with
its centre closer to the tails where the fogs converge, i.~e.~closer to the
observer. This is the case of the local super cluster which includes Virgo.
Fig.~\ref{fig:zoom} shows the location of our Virgo candidate (massive cluster
within the cyan circle) with respect to the observer (red cross) with both
schemes. We find that the position of the Virgo candidate is estimated to be
about $2.5 \,h^{-1}\rm{Mpc}$ closer to the observer with the CFOG scheme than
with the SFOG one (distances of $6.8 \,h^{-1}\rm{Mpc}$ and $9.3
\,h^{-1}\rm{Mpc}$ respectively for the ensemble means of $25$ simulations each).

The robustness of the RSD treatment can be seen in the uncertainty of the
position of particular objects. The spread in distance of the Virgo candidate
across all 25 ALPT+SFOG constrained simulations is very small (1 $\sigma$ $<$ 1
$h^{-1}$ Mpc), as shown in Fig.~\ref{fig:Virgo}. Also the mass of the
corresponding object is strongly constrained as shown in the right panel. This
indicates that a constrained environment leads to a constrained final mass of
halos. We plan to investigate this further and plan to study the mass evolution
of particular objects and the influence of the environment.

 It is possible to improve the SFOG scheme in various ways.  The description of
coherent velocities should be revised.  Furthermore, a radial dependent
dispersion velocity needs to be implemented to account for observational
selection effects, as we plan to do in future work. 
Nevertheless, the SFOG approach as presented in this work leads to constrained
simulations, which solve many of the problems present with the common CFOG
scheme  to great extent. For this reason we will restrict from now on our study
to the set of 25 ALPT+SFOG based constrained simulations.

\subsection{Comparison with the observations}

\label{sec:comp}

The cross power-spectra between the ALPT reconstructions with the SFOG treatment
and the galaxy field are shown on the left panel in Fig.~\ref{fig:XPS}. 
They present very high values up to $k\sim1\,h\,\rm{Mpc}^{-1}$ and are still
correlated at  $2.2<k<3\,h\,\rm{Mpc}^{-1}$ (1 $\sigma$ area of the zero
crossing of the cross power-spectra). The exact value of the zero crossing of
the cross-powerspectrum of a single simulation inhibits some stochasticity, due
to weak (anti-)correlations of random modes at the transition-scale from
constrained to random. Such incidents of matching phases are present in all
correlations of random fields. However by investigating the ensemble of
simulations we are able to quantify this random process and we are able to
specify this $ 68 \% $ C.I. \\
We select the 25 most correlated samples 
represented by the different coloured  lines as shown in Fig.~\ref{fig:XPS}.
Those samples, which are higher correlated with the galaxy field at a given $k$,
 show in most of the cases higher correlations in the entire $k$-space. Such a
rank-ordered behaviour is broken in some cases when approaching the Nyquist
frequency at $k\gsim2\,h\,\rm{Mpc}^{-1}$.  This demonstrates the robustness of
our selection approach focusing at $k=1\,h\,\rm{Mpc}^{-1}$ (see \S
\ref{sec:sel}). The required threshold at $k=1$ to select the 25 samples was:
$XP_{\rm th}=0.43$. By focusing only on the best correlated reconstructions we
do not sample the full parameter space of primordial density fluctuations, but
focus on the regions which are best correlated in the reconstructions.
The middle panel of Fig.~\ref{fig:XPS} shows the cross
power-spectra corresponding to the $N$-body simulations, being systematically
less correlated as compared to the ALPT ones. The reason for this is that the
initial conditions have 
essentially been computed within the ALPT approximation in addition to a
particular peculiar velocity model (2LPT with a dispersion term) to simulate
RSD. The $N$-body solutions  (modelling gravity more accurately) do not fully
agree with these models and introduce a small displacement. It would thus be
interesting to include in the future full gravity solvers within the
reconstruction process.   In the right panel we demonstrate that our RSD
treatment is over-all correct yielding higher correlations when transforming the
simulation from real-space (violet curve)  into redshift-space (cyan and red
curves). Here we consider two ways of transforming the data,  using the
peculiar velocities of the dark matter particles in one case (cyan) and  the
ones of the halo in the other (red). 
The right panel in Fig.~\ref{fig:XPS} demonstrates that our RSD modelling is
over-all correct as the cross-power spectra are clearly higher when including
redshift-space distortions. The lower correlation when using the halo velocity
fields is due to the additional stochastic process of collapsing objects along
the line-of-site. The larger dispersion when taking the DM velocities enhances
the probability that a number of simulated particles match the positions of
observed galaxies.  Nevertheless, matter is distributed in a more realistic
manner using the halo velocity fields as we will discuss below. We would like to
emphasise that we take the galaxy field according to Eq.~\ref{eq:deltaG}
including redshift-space distortions (also fogs) in our cross-correlation
analysis.
%  \citep[hence performing a direct comparison with the galaxy
% density field, where][for example, cross-correlate with a processed
% field]{Wang2013}.

\input{./fig_c2c}

To further assess the quality of the constrained simulations we make a
cell-to-cell comparison between the simulated and the observed density fields in
configuration space.
The correlation between two fields  $\phi_1$ and $\phi_2$ can be quantified
through the Pearson correlation coefficient $r$
% \be
%  r(\delta_1,\delta_2)=
% \frac{\sum_{i}\left(\delta_{1i}-\overline{\delta_1}\right)\left(\delta_{2i}
% -\overline{\delta_2}\right)}{\sqrt{
% \Big(\sum_{i}\left(\delta_{1i}-\overline{\delta_1}\right)\Big)^2
% \Big(\sum_{j}\left(\delta_{2j}-\overline{\delta_2}\right)\Big)^2 } } \,.
% \ee
\be
 r(\phi_1,\phi_2)=
\frac{\sum_{i}\left(\phi_{1i}-\overline{\phi_1}\right)\left(\phi_{2i}
-\overline{\phi_2}\right)}{\sqrt{
\Big(\sum_{i}\left(\phi_{1i}-\overline{\phi_1}\right)\Big)^2
\Big(\sum_{j}\left(\phi_{2j}-\overline{\phi_2}\right)\Big)^2 } } \,.
\ee
Let us focus here on the logarithmic overdensity $\phi\equiv\ln(1+\delta)$ for
various
reasons: first this representation yields an estimate of the linear density
field component \citep[see][]{Neyrinck2009,kitaura_lin}; second it suppresses
shot-noise in overdense areas\footnote{Let us write the galaxy overdensity field
as the sum of the expected galaxy density field and the shot-noise term:
$\delta_{\rm G}=\delta^{\rm exp}_{\rm G}+\epsilon$ \citep[see][]{kitaura_log}.
The residual shot-noise term is accordingly given in the logarithmic
representation by: $\ln(1+\delta_{\rm G})-\ln(1+\delta^{\rm exp}_{\rm
G})=\ln\left(1+\frac{\epsilon}{1+\delta^{\rm exp}_{\rm G}}\right)$.}. 
\input{./fig_maps}

\begin{figure*}
\begin{tabular}{cc}
\hspace{-3mm}
\includegraphics[width=0.47\textwidth]{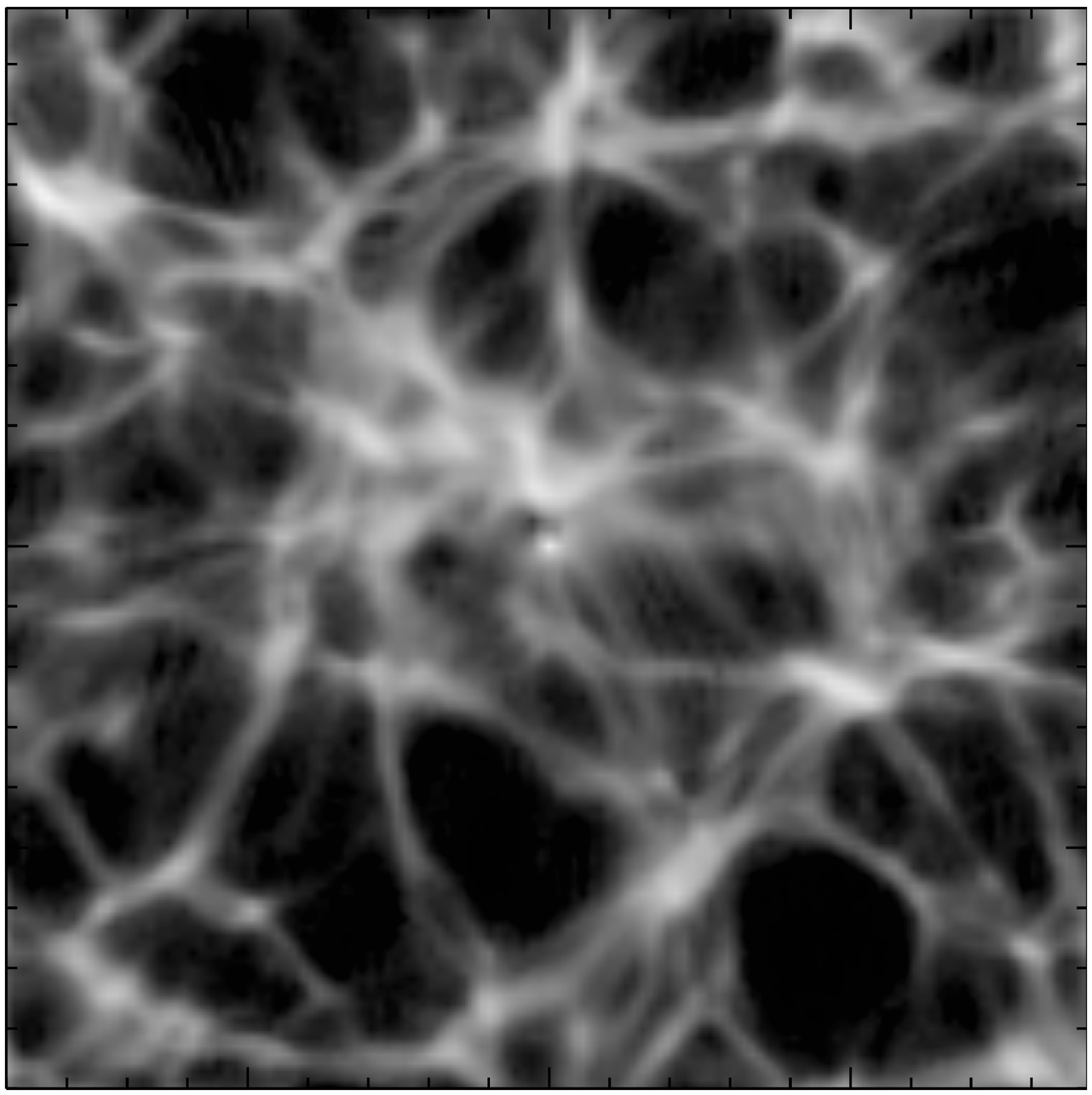}
\put(-232,130){\rotatebox[]{90}{SG Y [$h^{-1}$ Mpc]}}
\put(-214,190){$50$}
\put(-211,133){$0$}
\put(-217,78){$\shortminus50$}
\hspace{-0.05cm}
\includegraphics[width=0.47\textwidth]{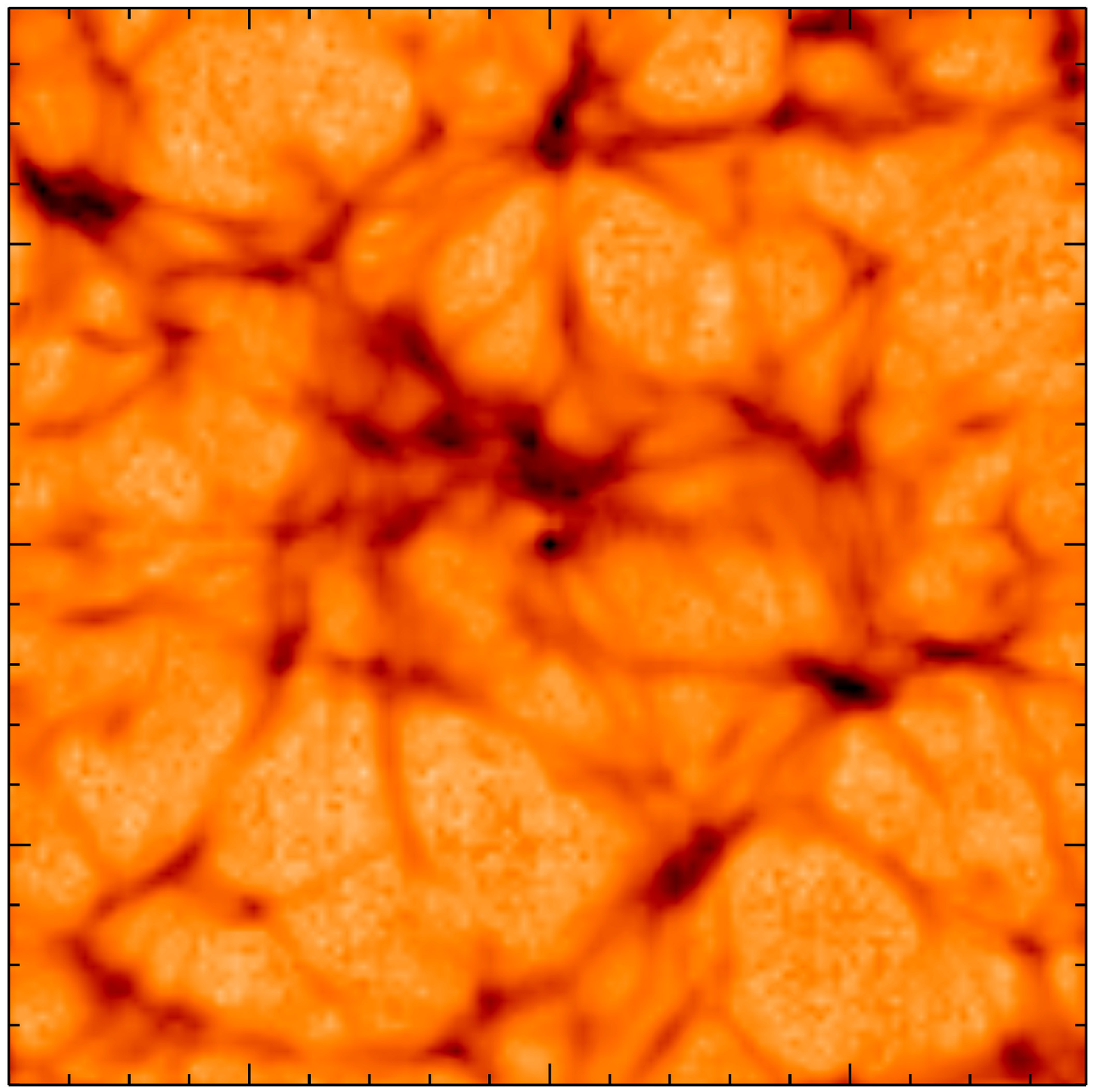}
\put(-12,130){\rotatebox[]{-90}{SG Y [$h^{-1}$ Mpc]}}
\put(-29,190){$ 50$}
\put(-28,133){$0$}
\put(-32,78){$\shortminus50$}
\vspace{-1.3cm}
\\
\hspace{-3mm}
\includegraphics[width=0.47\textwidth]{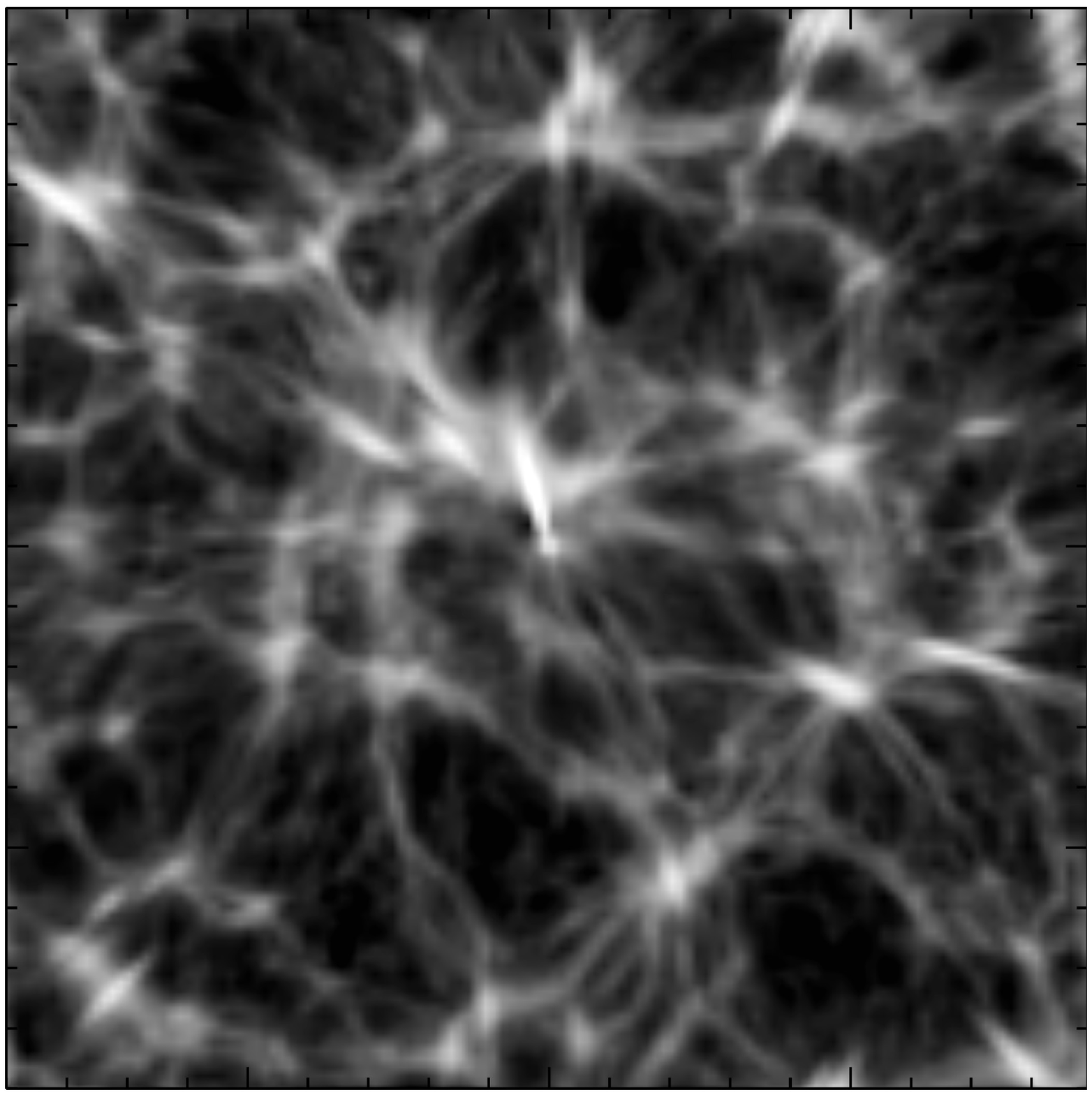}
\put(-232,130){\rotatebox[]{90}{SG Y [$h^{-1}$ Mpc]}}
\put(-122,15){\rotatebox[]{0}{SG X [$h^{-1}$ Mpc]}}
\put(-214,190){$50$}
\put(-211,133){$0$}
\put(-217,78){$\shortminus50$}
\put(-163,26){$\shortminus50$}
\put(-101.5,26){$0$}
\put(-47,26){$50$}
\hspace{-0.05cm}
\includegraphics[width=0.47\textwidth]{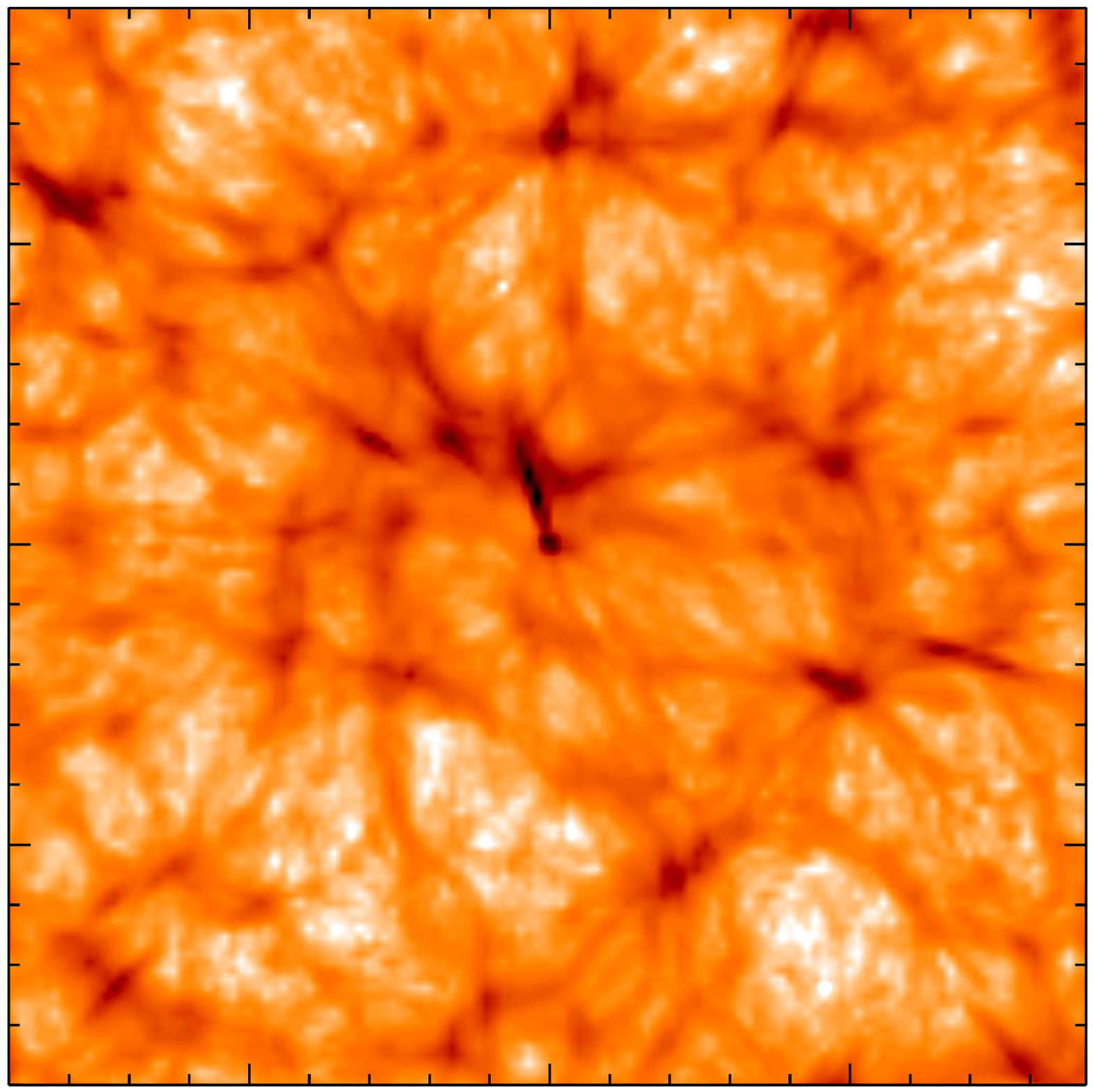}
\put(-158,15){\rotatebox[]{0}{SG X [$h^{-1}$ Mpc]}}
\put(-197,26){$\shortminus50$}
\put(-136.5,26){$0$}
\put(-83,26){$50$}
\put(-12,130){\rotatebox[]{-90}{SG Y [$h^{-1}$ Mpc]}}
\put(-29,190){$50$}
\put(-28,133){$0$}
\put(-32,78){$\shortminus50$}
\vspace{-5mm}
\end{tabular}
\caption{Signal and signal-to-noise maps in the supergalactic XY plane. {\bf Left panels:} Ensemble-averaged logarithm of the overdensity fields $\left<
\rm{log}\left( \rho/ \overline{\rho}\right)\right>$ of ALPT+SFOG reconstructions (top) and
constrained simulations (bottom) showing the middle slice of a $128^3$ CIC-grid.
{\bf Right panels:} Ensemble-averages divided by the standard deviation of the ensemble of reconstructions (top) and simulations (bottom) see Eq.~\ref{eq:SNR}.}
\label{fig:SNR_maps}
\end{figure*}

Fig.~\ref{fig:c2c} shows the average cell-to-cell correlation between the
$N$-body simulations and the galaxy field, and respectively ALPT,  in different
grid resolutions as compared to the ALPT reconstructions. We find that the
uncertainty is much larger in the underdense regions, as can be seen in the
increasing scatter towards negative values. The close alignment with the
45$^{\rm o}$ slope in the upper panels demonstrates that the  galaxy bias is
very close to unity, especially on scales larger than $r_{\rm S}=3.5\,h^{-1}$
Mpc. The deviations from the 45$^{\rm o}$ slope  shown in the lower panels are
due to the perturbative approach of the ALPT approximation, which does not fully
capture the nonlinear behaviour of gravitational clustering
\citep[see][]{Kitaura2012ALPT}.

The corresponding  cross-correlation coefficients are very high, being
$r=98.3\pm0.1$ \% on smoothing scales of $r_{\rm S}=3.5\,h^{-1}$ Mpc for both
the ALPT reconstructions and the constrained simulations. For $r_{\rm
S}=1.4\,h^{-1}$ Mpc we get $r=78.6\pm0.3$ \% and  $73.8\pm0.1$ \%, respectively.

Interestingly, if we keep the constrained simulations in real-space, we still
obtain $r=97.5\pm0.1\%$ with $r_{\rm S}=3.5\,h^{-1}$ Mpc for the
cross-correlation between the galaxy field (in redshift-space).
% It is remarkable that we still obtain $r=97.5\pm0.1$ \% with $r_{\rm
% S}=3.5\,h^{-1}$ Mpc  for the cross-correlation between the galaxy field (in
% redshift-space) and the constrained simulations in real-space.
This tells us
that small changes in $r$ of $\sim1\%$ with $r_{\rm S}=3.5\,h^{-1}$ Mpc can
imply improvements in the precision of the reconstruction of the order of
$\sim5\,h^{-1}$ Mpc (the order of coherent RSD). Such an estimation is important
when comparing with previous works \citep[see][where a value of $r=97\%$ is
obtained taking a smaller volume made by a sphere with a radius of 60 $h^{-1}$
Mpc and a volume limited 2MRS galaxy field using the same $r_{\rm
S}$]{Lavaux2010}.

Fig.~\ref{fig:maps} visualises the accuracy of the constrained simulation by
plotting on top of the dark matter density field the distribution of galaxies.
It is remarkable how accurately the red dots, each one standing for a galaxy,
match the simulated structures in redshift-space. The constrained simulations
are even able to connect filaments traced by a few galaxies. 
The upper panels show the maps using the DM velocities, while the middle ones
are based on the halo velocity field. For definiteness we move the simulation
particles according to the velocity of their host halo if available. For non
virialised particles we use the halo velocity field on a grid of $128^3$. We can
clearly see that in the upper panels fogs are overestimated.
Structure formation forms bound objects, therefore the correct way of modelling
RSD consists of taking the halo velocity field to transform the simulation from
real-space to redshift-space. To accurately model the RSD one needs a high
enough resolution to resolve the masses of the observed objects and consider
selection effects. The velocity dispersion of galaxies decreases with distance
to the observer, since magnitude cuts tend to select more massive objects at
increasing distances.  We plan to perform a study of high  resolution $N$-body
simulations to study this in detail. Our current simulations yield already
realistic fogs, as can be seen from the middle panels in Fig.~\ref{fig:maps}.
The lower panels in that figure show a {\it natural} real-space matter
distribution with more extended structures along the line-of-site as opposed to
the squashed structures in redshift-space (see Coma, Great Attractor and
Perseus-Pisces region). Here we also find that the Local Void is less empty than
it appears in redshift-space.

\subsection{Statistical analysis}

\label{sec:stats}

Here we want to investigate the robustness of the constrained simulations by
looking at their statistical properties. 
Fig.~\ref{fig:SNR_maps} shows in the left panels the ensemble-averaged densities
in logarithmic space $\left<\rm{log}(\delta +1)\right> = \left<\rm{log}\left(
\rho/ \overline{\rho}\right)\right>$ corresponding to the set of ALPT
reconstructions and simulations. This demonstrates the consistency between the
ALPT and the $N$-body simulations on scales larger than a few Mpc confirming our
cell-to-cell comparison. However the $N$-body simulations show as expected,
much more small-scale power. Looking at the signal-to-noise ratio:
\be
  \frac{S}{N} \equiv \frac{{\left\langle\rm{log}\left(1+\delta_{\rm
DM}\right)\right\rangle}}{\sqrt{\left\langle\left(\rm{log}\left(1+\delta_{\rm
DM}\right)\right)^2 - \left\langle\rm{log}\left(1+\delta_{\rm
DM}\right)\right\rangle^2 \right\rangle}}\,,
\label{eq:SNR}
\ee
where the ensemble mean is computed by volume averaging, we find that the
uncertainty is smaller (darker colours) for the ALPT case. The small-scale power
in the $N$-body case can be especially seen  in the  larger  uncertainties in
low density regions.
\input{./fig_pdfm}
To accurately assess the statistical quality of the constrained simulations we
construct a reference sample of 25 randomly seeded $N$-body simulations with the
same setup.

\begin{figure}
\vspace{-.7cm}
\begin{center}
\hspace{-1.cm}
\includegraphics[width=.4\textwidth]{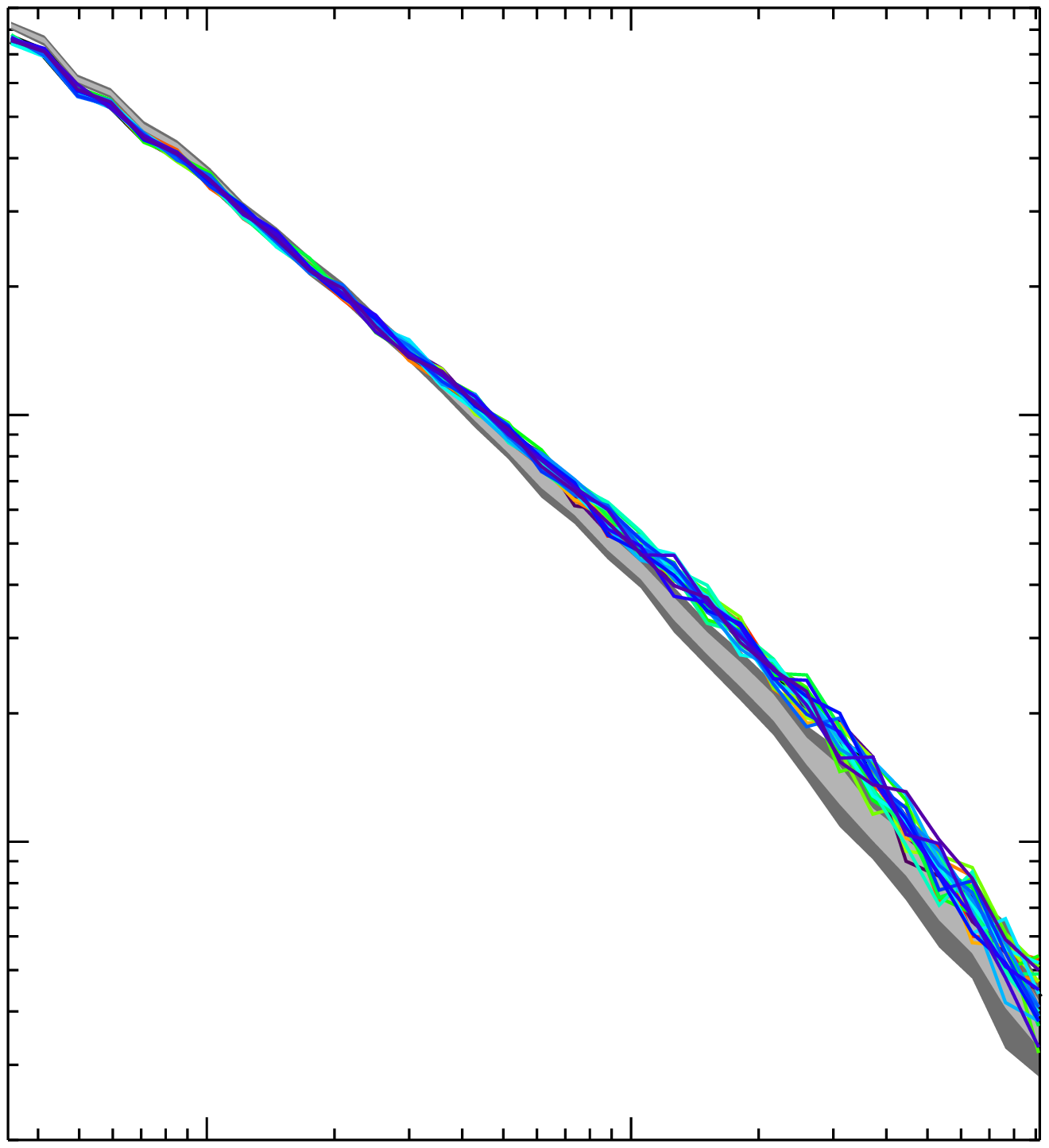}
\put(-175,185){$10^4$}
\put(-175,128){$10^3$}
\put(-175,68){$10^2$}
\put(-139,15){$10^{12}$}
\put(-77,15){$10^{13}$}
\put(-190,110){\rotatebox[]{90}{${\rm d}N/{\rm d}\;{\rm ln}M$}}
\put(-110,5){$M_{200} [h^{-1}M_{\odot}]$}
\caption{Mass function $N(M)$ of halos in the ensemble of $25$ constrained
simulations (same colour-code as Fig.~\ref{fig:XPS}). The light and dark grey
shaded contours indicate 1 and 2 sigma regions according to the 25 random
simulations.}
\label{massF}
\end{center}
\end{figure}

\begin{figure}
\begin{center}
\vspace{-2mm}
\hspace{-6mm}
\includegraphics[width=0.5\textwidth]{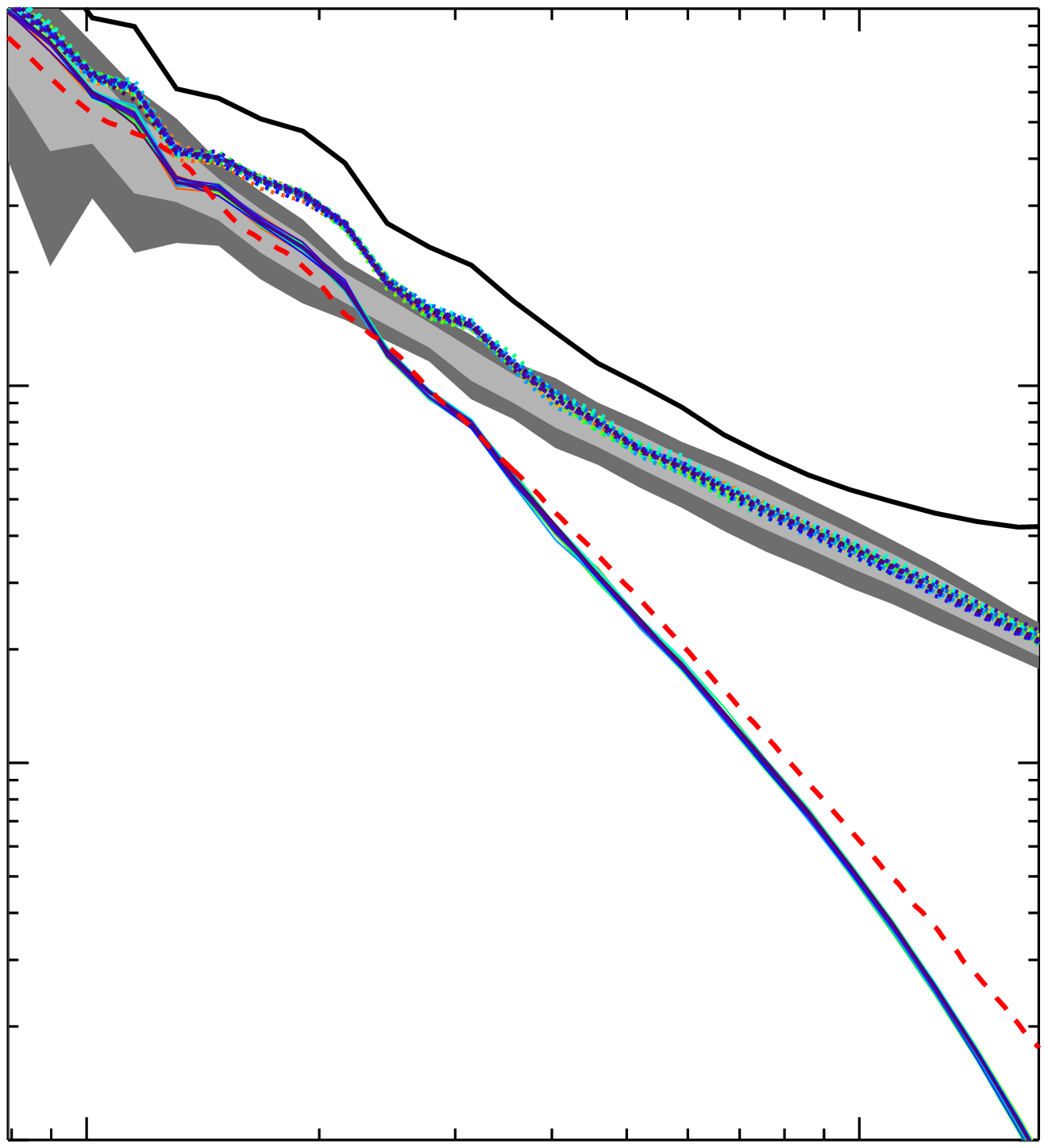}
% _Pk_withresi_180_0384}
\put(-215,164){$10^3$}
\put(-215,98){$10^2$}
\put(-215,32){$10$}
% \put(-233,149){$10.0$}
% \put(-230,103){$1.0$}
% \put(-230,54){$0.1$}
\put(-192,22){$0.1$}
\put(-56,22){$1.0$}
\put(-177.5,160){$P_{\rm G}^z(k)$}
\put(-190,163){\line(1,0){9}}
\put(-177.5,140){\color{blue} $P_{N{\rm body}}^z(k)$}
% \put(-190,223){\color{blue}\line(1,0){9}}
\put(-177.5,120){\color{red} $P_{\rm lin}(k)$}
\put(-190,123){\color{red}\line(1,0){4}}
\put(-184,123){\color{red}\line(1,0){4}}
% \put(-190,180){\color{blue} $\cdots\,P_{\rm ALPT}^z(k)$}
\put(-177.5,100){\color{blue} $P_{\rm ALPT}^z(k)$}
\put(-190,140){\color{blue} $\cdots\,$}
\put(-190,103){\color{blue}\line(1,0){9}}
\put(-245,150){\rotatebox[]{90}{\rm $P(k) $}}
% \put(-245,106){\rotatebox[]{90}{\rm $P(k)/\langle P^z_{\rm RS}(k)\rangle$ 222}}
\put(-135,9){\rm $k$ [$h$ Mpc$^{-1}$]}
\hspace{-6mm}
\vspace{-4mm}
\caption{Power-spectra of the 2MRS galaxy overdensity (black) the ALPT+SFOG
reconstructions in real sapce (coloured solid), and the constrained simulations
in real space (coloured dotted). The light and dark grey shaded contours
indicate 1 and 2 sigma regions according to the 25 random simulations.}
\label{fig:ps}
\end{center}
\end{figure}

We checked that the white-noise fields used for the initial conditions have
numerically vanishing statistical moments: mean, skewness and kurtosis.
Fig.~\ref{fig:IC_PDF} shows the matter statistics for the constrained
simulations at redshifts 100 and 0. We find that the curves lie within the
variance of the reference random simulations.
The mass function (Fig.~\ref{massF}) shows a slight overproduction of
massive haloes for the mass range $M_{200}>10^{13}\,h^{-1}M_{\odot}$, being,
however, compatible with the reference set within the error bars. It is unclear,
whether the particular realisation of the observed volume could be responsible
for this issue.

Finally, we analyse the power-spectra to verify the method and the quality of
the constrained simulations. 
This is shown in Fig.~\ref{fig:ps}. Here one can see that the specific features
of the galaxy power-spectrum, characteristic of the particular matter
distribution, are preserved both in the ALPT reconstructions and in the
constrained simulations. We can also see that the CSs have more power than the
ALPT reconstructions, correcting for the bias caused by the perturbative
structure formation approach.
 The black solid line shows the excess of power for the galaxy field due to the
galaxy bias and shot-noise. The solid lines show the strongly biased nature of
the ALPT+SFOG model. The constrained simulations however, are close to
unbiased with respect to the random simulations and show the characteristic
features of the particular realisation of the Local Universe. 
Fig.~\ref{fig:ps} demonstrates that we are reasonably modeling, in a scale
dependent way, the biased nature of the galaxy field and the nonlinear
gravitational clustering up to high $k$s ($k\gsim2\,h\,{\rm Mpc}^{-1}$).

% \vspace{2cm} %%% to push it to new site

\section{Conclusions}

We have presented in this work constrained $N$-body simulations of the Local
Universe based on the 2MRS galaxy redshift catalogue. This is the first time that
a statistical self-consistent {\it forward} approach is applied to
observations to
obtain the fully nonlinear phase-space distribution. Our approach is based on
sampling the posterior distribution function of a Gaussian prior with a
likelihood relating the initial fields with the observations through a
particular structure formation model. In this way we can exploit the high number
density of observed galaxies in the Local Universe and use between 2 and 3
orders of magnitude more constraints for the initial fields than previous works.
% % The high number of
% % about 2 Million constraints for the initial field represent information of
% % observed regions with detected galaxies as well as those with no detections.
Special care has been taken to
include recent improvements in modelling structure formation and redshift-space
distortions within the reconstruction method of the primordial fluctuations.
In particular, Augmented Lagrangian Perturbation Theory (ALPT) was used and
redshift-space distortions were modelled, hence taking into account the tidal
shear tensor and including the modelling of virialised motions.

Our studies have
shown, that this effort is necessary to deal with such a large number of
constraints, while resolving the density field down to scales of 
 2-3 $\,h^{-1}\rm{Mpc}$. Otherwise, strong artifacts are introduced in
the constrained simulations due to transients which originated through
shell-crossing in the iterative reconstruction process.
We verified that this is the case, when using standard second order Lagrangian
perturbation theory or when using the galaxy distribution where the
fingers-of-god have been collapsed. Parallel filamentary structures
perpendicular to the line-of-sight and negative fingers-of-god in real-space
appear in such cases. We also found that the position of clusters which are
distorted as fingers-of-god close to the observer (such as Virgo) tend to be
systematically closer in the collapsing schemes. We have managed to strongly
suppress all these effects with our improved treatments. Nevertheless, we
consider that further investigation should be carried out improving the models
with respect to the statistical properties of the constrained simulations.
One focus should be on the mass functions. We found, that those show a slight
overproduction of massive haloes, which may or may not be induced by an
artifact. This is not clear, as the deviation is small and can be considered
compatible with the random simulations. Especially so since one must also
consider that the Local Universe is a particular realisation, which could have
very specific statistical characteristics.
Nevertheless the compatibility between the constrained simulations and the
random ones in the matter statistics at starting and final redshifts and in the
power-spectra, demonstrates the quality of the constrained simulations.
Certainly, we should improve our selection function treatment, including
a real-space estimation to treat the Kaiser-rocket effect. We have considered
that this is negligible given the small volume we are studying. In any case,
such an improvement would also permit us to tackle larger volumes for which the
selection function drops more dramatically. In those studies one could also
investigate the peculiar motions on larger scales, including more distant
attractors, since now we are limited by the size of the box
($180\,h^{-1}\,\rm{Mpc}$) and the periodic boundary conditions.

This work serves as a consistency check for our understanding of structure
formation. In this sense the high level of resemblance between our models and
the actual galaxy distribution corroborate the initial assumptions and therefore
do not indicate any deviations from the standard cosmological model. It would be
interesting to extend this kind of study varying the assumptions to further test
our models and verify whether these are the most compatible with observations.
The level of precision that we achieve already in this work should help to get
new insights into fields of study that require accurate density estimates of the
Local Universe such as cosmic rays, the warm hot inter-galactic medium or
Dark Matter annihilation and decay signals.
The full nonlinear velocity field determined in this work can contribute to our
understanding of the cosmic flows in the Local Universe. Furthermore the
simulations can also provide a consistent formation history of particular
clusters.
We plan to exploit this by studying objects within their correct environment,
explore their formation histories and compare them with observations.

\section*{Acknowledgements}

The authors thank the CLUES collaboration\footnote{www.clues-project.org}, in
particular Yehuda Hoffman, Gustavo Yepes and Timur Doumler for very encouraging
and helpful discussions. 
Special thanks to Arman Khalatyan for providing us with the friends-of-friends
code, Steffen Knollmann for assistance with \textsc{ginnungagap}, Pirin Erdo{\u
g}du for providing us the augmented 2MRS catalogue.
% and A.~Knebe for the publicly available amiga halo finder \textsc{ahf}. 
SH acknowledges support by the Deutsche Forschungsgemeinschaft under the grant
$\rm{GO}563/21-1$.

\bibliography{paper}
\bibliographystyle{mn2e.bst}

\label{lastpage}
\end{document}